\documentclass[10pt]{article}
\usepackage{graphicx}
\usepackage{amsmath}
\usepackage{amssymb}
\usepackage{balance}
\usepackage{bm}
\usepackage{caption2}
\usepackage{indentfirst}

\begin{document}
\bibliographystyle{prsty}
\begin{center}
{\large {\bf \sc{ Analysis of the possible $D\bar{D}_{s0}^*(2317)$ and $D^*\bar{D}_{s1}^*(2460)$ molecules with QCD sum rules }}} \\[2mm]
Zun-Yan Di$^{1,2}$, Zhi-Gang Wang$^{1}$\footnote{E-mail: zgwang@aliyun.com. }, Guo-Liang Yu$^{1}$    \\
$^{1}$ Department of Physics, North China Electric Power University, Baoding 071003, P. R. China \\
$^{2}$ School of Nuclear Science and Engineering, North China Electric Power University, Beijing 102206, P. R. China
\end{center}

\begin{abstract}
In this article, we assume that there exist the pseudoscalar $D\bar{D}_{s0}^*(2317)$ and $D^*\bar{D}_{s1}^*(2460)$ molecular states $Z_{1,2}$
and construct the color singlet-singlet molecule-type interpolating currents
to study their masses with the QCD sum rules.
In calculations, we consider the contributions of the vacuum condensates up to dimension-10
and use the formula $\mu=\sqrt{M_{X/Y/Z}^{2}-\left(2{\mathbb{M}}_{c}\right)^{2}}$
to determine the energy scales of the QCD spectral densities.
The numerical results,
$M_{Z_1}=4.61_{-0.08}^{+0.11}\,\text{GeV}$
and $M_{Z_2}=4.60_{-0.06}^{+0.07}\,\text{GeV}$,
which lie above the $D\bar{D}_{s0}^*(2317)$ and $D^*\bar{D}_{s1}^*(2460)$ thresholds respectively,
indicate that the $D\bar{D}_{s0}^*(2317)$ and $D^*\bar{D}_{s1}^*(2460)$
are difficult to form   bound state molecular states, the $Z_{1,2}$ are probably resonance states.
\end{abstract}



\section{Introduction}
In the recent years, many new charmonium-like and bottomonium-like exotic mesons \cite{meson}
(being bosons like the traditional $q\bar{q}$ mesons),
have been observed experimentally,
and are labeled as the $XYZ$ states \cite{XYZ}.
These exotic states,
with growing evidences, cannot be the pure $c\bar{c}$ or $b\bar{b}$ states and
are considered as good candidates for tetraquark states,
which do not fit into the conventional quark model picture \cite{quark-model}.
In order to decipher their underlying structure,
a number of interpretations have been proposed, such as
the molecules \cite{molecule1,molecule}, the tetraquark states \cite{tetraquark},
the hybrid mesons \cite{Hybrid},
the kinematical effects \cite{kinematical-effect}, and so on.

In the molecular picture,
a four-quark state is explained as a weakly bound state of two mesons \cite{molecule-model}.
Each constituent   meson is bound internally by strong QCD color
forces, while the mesons bind to
each other by means of a much weaker color-neutral residual QCD force,  analogous   of the van der Waals attraction in chemistry.
Among these observed $XYZ$ states,
some lie remarkably close to the meson-meson thresholds.
Therefore, the molecular interpretation seems plausible for these states.
The most impressive example is the original exotic state,
the $X(3872)$ \cite{X(3872)},
which has been investigated as the $D\bar{D}^*$ molecular state
by many theoretical groups \cite{molecule1,groups},
owing to its mass with
$m_{X(3872)}-m_{D^{*0}}-m_{D^{0}}=+0.01\pm0.18\, \text{MeV}$.
The $Z_c(3900)$, observed  by the BESIII collaboration firstly in 2013 \cite{BESIII},
is also close to the threshold of $D\bar{D}^*$,
and is taken as the isovector partner of
the established  isoscalar bound molecular state $X(3872)$
with the same quantum number $J^{P}=1^{+}$ in some references \cite{Z(3872),Z(3900)}.
Interestingly,
the observed bottomonium-like states $Z_b(10610)$ and $Z_b(10650)$ by the Belle collaboration \cite{Belle},
own the same near-threshold nature
and are interpreted successfully as the $B\bar{B}^*$ and $B^*\bar{B}^*$ molecular states \cite{Zb}.
The successes of the molecular interpretation for some observed exotic states
stimulate the further theoretical studies on the  analogous open-charmed meson pair system
as a bound molecular state,
which make several predictions of the possible molecules.

In theoretical techniques,
the QCD sum rules method is a powerful tool in studying the hidden-charm (bottom)
tetraquark or molecular states and hidden-charm pentaquark states.
Here,
we make the assumption that there exist
the pseudoscalar $D\bar{D}_{s0}^*(2317)$ and $D^*\bar{D}_{s1}^*(2460)$ molecular states,
and study their masses with the QCD sum rules to check the existence of the corresponding molecular states.
The $D$ and $D^*$ mesons have negative parity,
while the parity is positive for the $\bar{D}_{s0}^*(2317)$ and $\bar{D}_{s1}^*(2460)$ mesons.
Based on the theoretical analysis in Ref. \cite{PRD97-091501},
there is a Coulomb-like force by exchanging a kaon in S-wave, that might be able to
bind the $D\bar{D}_{s0}^*(2317)$ and $D^*\bar{D}_{s1}^*(2460)$ systems, respectively.
In addition,
the mass difference between the $D$($D^*$) and $\bar{D}_{s0}^*$($\bar{D}_{s1}^*$)
is close to the kaon mass,
which means that the exchanged kaon will be near the mass shell
and the range of the interaction will be unusually large.
These are the reasons why
we are interested in the $D\bar{D}_{s0}^*(2317)$ and $D^*\bar{D}_{s1}^*(2460)$ molecules.
In calculations, we consider the contributions of the vacuum condensates up to dimension-10,
and use the formula $\mu=\sqrt{M_{X/Y/Z}^{2}-\left(2{\mathbb{M}}_{c}\right)^{2}}$
to determine the energy scales of the QCD spectral densities \cite{Z(3872)},
which can enhance the pole contributions remarkably and improve the convergent behaviors
of the operator product expansion in the QCD sum rules for the exotic hadrons \cite{parameters}.

The rest of this article is arranged as follows.
In section 2,
we consider the $D\bar{D}_{s0}^*(2317)$ and $D^*\bar{D}_{s1}^*(2460)$ systems as the pseudoscalar molecules,
construct the corresponding color singlet-singlet molecule-type interpolating currents,
and extract their masses and pole residues with the QCD sum rules.
The numerical results and discussions are performed in section 3.
The last section is reserved for our conclusion.

\section{QCD sum rules for the possible $D\bar{D}_{s0}^*(2317)$ and $D^*\bar{D}_{s1}^*(2460)$ molecular  states}
Based on our assumption that there exist the pseudoscalar $D\bar{D}_{s0}^*(2317)$ and $D^*\bar{D}_{s1}^*(2460)$ molecular states,
the corresponding color singlet-singlet molecule-type interpolating currents are written as

\begin{eqnarray}\label{eq1}
J_{1}(x)&=&\bar{q}^a(x)i\gamma_5 c^a(x)\bar{c}^b(x)s^b(x)\,,
\end{eqnarray}
and
\begin{eqnarray}\label{eq2}
J_{2}(x)&=&\bar{q}^a(x)\gamma_\mu c^a(x)\bar{c}^b(x)\gamma^\mu\gamma_5 s^b(x)\,,
\end{eqnarray}
respectively,
where $a$, $b$ are color indexes,
and $q$ denotes an up or down quark.

In QCD sum rules, we consider the two-point correlation functions
\begin{eqnarray}\label{correlation-function}
\Pi_{1,2}\left(p\right)&=&i\int d^{4} x e^{ip\cdot x}
\langle0|T\left\{J_{1,2}(x) J_{1,2}^{\dag}(0)\right\}|0\rangle\, ,
\end{eqnarray}
which can be obtained  in two ways:
on the phenomenological side and at the quark level.

On the phenomenological side, we insert a complete set of intermediate hadronic states with the
same quantum numbers as the current operators $J_{1,2}(x)$
into the correlation functions $\Pi_{1,2}\left(p\right)$
to obtain the hadronic representations.
After isolating the ground state contributions from the pole terms,
we get the following results,
\begin{eqnarray}\label{correlation-function-1}
\Pi_{1,2}\left(p\right)
&=&\frac{\lambda_{Z_{1,2}}^2}{M_{Z_{1,2}}^2-p^2}+\cdots\, ,
\end{eqnarray}
where the $Z_1$, $Z_2$ denote the ground states, provisionally,
considered as the $D\bar{D}_{s0}^*(2317)$, $D^*\bar{D}_{s1}^*(2460)$ molecules,
respectively,
and the pole residues $\lambda_{Z_{1,2}}$ are defined by $\langle0|J_{1,2}(0)|Z_{1,2}(p)\rangle=\lambda_{Z_{1,2}}$,
which show the couplings of the currents $J_{1,2}$ to the states $Z_{1,2}$.

At the quark level,
we calculate the two-point correlation functions $\Pi_{1,2}\left(p\right)$
via the operator product expansion method in perturbative QCD.
We contract the $q$, $s$ and $c$ quark fields
with the wick theorem and obtain the following results:
\begin{eqnarray}
\Pi_{1}\left(p\right)&=&i\int d^{4}x e^{ip\cdot x}
{\rm Tr}\left[i\gamma_5 C^{aa'}(x)i\gamma_5Q^{a'a}(-x)\right]
{\rm Tr}\left[C^{b'b}(-x)S^{bb'}(x)\right] \,,\nonumber\\
\Pi_{2}\left(p\right)&=&i\int d^{4}x e^{ip\cdot x}
{\rm Tr}\left[\gamma_\mu C^{aa'}(x)\gamma_\nu Q^{a'a}(-x)\right]
{\rm Tr}\left[\gamma^\nu\gamma_5C^{b'b}(-x)\gamma^\mu\gamma_5S^{bb'}(x)\right] \,.
\end{eqnarray}
where the $Q_{ab}(x)$, $S_{ab}(x)$ and $C_{ab}(x)$ are the full $q$, $s$ and $c$ quark propagators, respectively,
\begin{eqnarray}
Q_{a b}(x)&=&\frac{i\delta_{ab}x\!\!\!/}{2\pi^{2}x^{4}}-\frac{\delta_{ab}\langle\bar{q}q\rangle}{12}
-\frac{\delta_{ab}x^{2}\langle\bar{q}g_{s}\sigma Gq\rangle}{192}-\frac{\delta_{ab}x^{2}x\!\!\!/g_{s}^{2}\langle\bar{q}q\rangle^{2}}{7776}
-\frac{i g_{s}G_{\alpha\beta}^{n}t_{ab}^{n}(x\!\!\!/\sigma^{\alpha\beta}+\sigma^{\alpha\beta}x\!\!\!/)}{32\pi^{2}x^{2}} \nonumber\\
&&-\frac{\delta_{ab}x^{4}\langle\bar{q}q\rangle\langle GG\rangle}{27648}
-\frac{1}{8}\langle\bar{q}_{b}\sigma^{\alpha\beta}q_{a}\rangle\sigma_{\alpha\beta}
-\frac{1}{4}\langle\bar{q}_{b}\gamma_{\mu}q_{a}\rangle\gamma^{\mu}+\cdots \ ,\label{propagator-u,d}\\
S_{a b}(x)&=&\frac{i\delta_{ab}x\!\!\!/}{2\pi^{2}x^{4}}-\frac{\delta_{ab}m_{s}}{4\pi^{2}x^{2}}
-\frac{\delta_{ab}\langle\bar{s}s\rangle}{12}+\frac{i\delta_{ab}x\!\!\!/m_{s}\langle\bar{s}s\rangle}{48}
-\frac{\delta_{ab}x^{2}\langle\bar{s}g_{s}\sigma Gs\rangle}{192} \nonumber\\
&&+\frac{i\delta_{ab}x^{2}x\!\!\!/m_{s}\langle\bar{s}g_{s}\sigma Gs\rangle}{1152}
-\frac{\delta_{ab}x^{2}x\!\!\!/g_{s}^{2}\langle\bar{s}s\rangle^{2}}{7776}
-\frac{i g_{s}G_{\alpha\beta}^{n}t_{ab}^{n}(x\!\!\!/\sigma^{\alpha\beta}+\sigma^{\alpha\beta}x\!\!\!/)}{32\pi^{2}x^{2}} \nonumber\\
&&-\frac{\delta_{ab}x^{4}\langle\bar{s}s\rangle\langle GG\rangle}{27648}
-\frac{1}{8}\langle\bar{s}_{b}\sigma^{\alpha\beta}s_{a}\rangle\sigma_{\alpha\beta}
-\frac{1}{4}\langle\bar{s}_{b}\gamma_{\mu}s_{a}\rangle\gamma^{\mu}+\cdots \ ,\label{propagator-s}\\
C_{a b}(x)&=&\frac{i}{(2\pi)^4}\int d^4 ke^{-ik\cdot x}\bigg\{\frac{k\!\!\!/ +m_{c}}{k^{2}-m_{c}^{2}}\delta_{ab}
-g_{s}t_{ab}^{n}G_{\alpha\beta}^{n}\frac{(k\!\!\!/+m_{c})\sigma^{\alpha\beta}+\sigma^{\alpha\beta}(k\!\!\!/+m_{c})}{4(k^{2}-m_{c}^{2})^{2}} \nonumber\\
&&+\frac{g_{s}t_{ab}^{n}D_{\alpha}G_{\beta\lambda}^{n}(f^{\lambda\alpha\beta}+f^{\lambda\beta\alpha})}{3(k^{2}-m_{c}^{2})^{4}} \nonumber\\
&&-\frac{g_{s}^{2}(t^{n}t^{m})_{a b}G_{\alpha\beta}^{n}G_{\mu\nu}^{n}(f^{\alpha\beta\mu\nu}+f^{\alpha\mu\beta\nu}+f^{\alpha\mu\nu\beta})}{4(k^{2}-m_{c}^2)^{5}}+\cdots\bigg\} \ ,
\end{eqnarray}
\begin{eqnarray}
f^{\lambda\alpha\beta}&=&(k\!\!\!/+m_{c})\gamma^{\lambda}(k\!\!\!/+m_{c})\gamma^{\alpha}(k\!\!\!/+m_{c})\gamma^{\beta}(k\!\!\!/+m_{c})\ ,\nonumber\\
f^{\alpha\beta\mu\nu}&=&(k\!\!\!/+m_{c})\gamma^{\alpha}(k\!\!\!/+m_{c})\gamma^{\beta}(k\!\!\!/+m_{c})\gamma^{\mu}(k\!\!\!/+m_{c})\gamma^{\nu}(k\!\!\!/+m_{c})\ ,
\end{eqnarray}
$t^{n}=\frac{\lambda^{n}}{2}$,
the $\lambda^{n}$ is the Gell-Mann matrix,
and $D_{\alpha}=\partial_\alpha-ig_{s}G_{\alpha}^{n}t^{n}$ \cite{PRT85}.
Then we compute the integrals in the coordinate space for the light quark propagator
and in momentum space for the charm quark part.
In the operator product expansion,
we take into account the contributions of vacuum condensates up to dimension-10,
assume vacuum saturation for the higher dimensional vacuum condensates,
and keep terms which are linear in the strange quark mass $m_s$.
The vacuum condensates are the vacuum expectations of the operators $\mathcal{O}_n( \alpha_s^{k})$.
We take the truncations $n\leq10$ and $k\leq1$ for the operators in a consistent way,
and discard the perturbative corrections.
In Eqs. \eqref{propagator-u,d}--\eqref{propagator-s},
we retain the terms $\langle\bar{q}_{b}\sigma_{\mu\nu}q_{a}\rangle$, $\langle\bar{s}_{b}\sigma_{\mu\nu}s_{a}\rangle$,
$\langle\bar{q}_{b}\gamma_{\mu}q_{a}\rangle$ and $\langle\bar{s}_{b}\gamma_{\mu}s_{a}\rangle$
originate from the Fierz re-arrangement of the $\langle q_{a}\bar{q}_{b}\rangle$ and $\langle s_{a}\bar{s}_{b}\rangle$
to absorb the gluons emitted from the heavy quark lines
so as to extract the mixed condensates and four-quark condensates
$\langle\bar{q}g_{s}\sigma Gq\rangle$,
$\langle\bar{s}g_{s}\sigma Gs\rangle$, $g_{s}^{2}\langle\bar{q}q\rangle^{2}$ and
$g_{s}^{2}\langle\bar{s}s\rangle^{2}$, respectively.
One can consult Ref. \cite{axialvector}
for some technical details about the operator product expansion.
Once the analytical expressions of the correlation functions $\Pi_{1,2}(p)$ are obtained,
the QCD spectral densities $\rho_{1,2}(s)$ are given by the imaginary parts of the correlation
functions: $\rho_{1,2}(s)=\frac{\text{Im}\Pi_{1,2}(s)}{\pi}$.

According to the quark-hadron duality,
we match the correlation functions $\Pi_{1,2}(p)$
obtained on the phenomenological side and at the quark level below the continuum thresholds $s_0$,
and perform Borel transform with respect to the variable $P^2 =-p^2$
to obtain the following QCD sum rules:
\begin{eqnarray}\label{PoleResidue}
\lambda_{Z_{1,2}}^{2}\exp\left(-\frac{M_{Z_{1,2}}^{2}}{T^{2}}\right)
&=&\int_{4m_{c}^{2}}^{s_{0}}ds\rho_{1,2}\left(s\right)\exp\left(-\frac{s}{T^{2}}\right)\ ,
\end{eqnarray}
where
\begin{eqnarray}
\rho_{1,2}\left(s\right)&=& \rho^{0}_{1,2}\left(s\right)+\rho^{3}_{1,2}\left(s\right)+\rho^{4}_{1,2}\left(s\right)
+\rho^{5}_{1,2}\left(s\right)+\rho^{6}_{1,2}\left(s\right)+\rho^{7}_{1,2}\left(s\right) \nonumber\\
&&+\rho^{8}_{1,2}\left(s\right)+\rho^{10}_{1,2}\left(s\right)\ ,
\end{eqnarray}
the superscripts 0, 3, 4, 5, 6, 7, 8, 10 denote the dimensions of the vacuum condensates,
and the $T^2$ denotes the Borel parameter.
The explicit expressions of the spectral densities $\rho_{1,2}(s)$ are collected
in the appendix.

To extract the masses of the states $Z_{1,2}$,
we take the derivative of Eq. \eqref{PoleResidue} with respect to $\frac{1}{T^2}$
and eliminate the pole residues $\lambda_{Z_{1,2}}$:
\begin{eqnarray}\label{mass}
M_{Z_{1,2}}^{2}&=&\frac{\int_{4m_{c}^{2}}^{s_{0}}ds\frac{d}{d\left(-1/T^{2}\right)}\rho_{1,2}\left(s\right)\exp\left(-\frac{s}{T^{2}}\right)} {\int_{4m_{c}^{2}}^{s_{0}}ds\rho_{1,2}\left(s\right)\exp\left(-\frac{s}{T^{2}}\right)}\ .
\end{eqnarray}

\section{Numerical results and discussions}
In this section, we perform the numerical analysis.
To extract the numerical values of $M_{Z_{1,2}}$,
we take the standard values of the vacuum condensates
$\langle\bar{q}q\rangle=-(0.24\pm0.01\,\text{GeV})^{3}$, $\langle\bar{s}s\rangle=(0.8\pm0.1)\langle\bar{q}q\rangle$,
$\langle\bar{q}g_{s}\sigma Gq\rangle=m_{0}^{2}\langle\bar{q}q\rangle$,
$\langle\bar{s}g_{s}\sigma Gs\rangle=m_{0}^{2}\langle\bar{s}s\rangle$,
$m_{0}^{2}=(0.8\pm0.1)\,\text{GeV}^{2}$, $\langle\frac{\alpha_{s}GG}{\pi}\rangle=(0.33\,\text{GeV})^{4}$
at the energy scale  $\mu=1\,\text{GeV}$ \cite{PRT85,SVZ79,ColangeloReview},
choose the $\overline{\text{MS}}$ masses $m_{c}(m_c)=(1.28\pm0.03)\,\rm{GeV}$,
$m_{s}(\mu=2\,\rm{GeV})=(0.096^{+0.008}_{-0.004})\,\rm{GeV}$ from the Particle Data Group \cite{XYZ},
and neglect the up and down quark masses, i.e., $m_u=m_d=0$.
Moreover, we take into account the energy-scale dependence of the input parameters
on the QCD side from the renormalization group equation,
\begin{eqnarray}
\langle\bar{q}q\rangle(\mu)&=&\langle\bar{q}q\rangle(Q)\left[\frac{\alpha_{s}(Q)}{\alpha_{s}(\mu)}\right]^{\frac{4}{9}}\, ,\nonumber\\
\langle\bar{s}s\rangle(\mu)&=&\langle\bar{s}s\rangle(Q)\left[\frac{\alpha_{s}(Q)}{\alpha_{s}(\mu)}\right]^{\frac{4}{9}}\, ,\nonumber\\
\langle\bar{q}g_{s}\sigma Gq\rangle(\mu)&=&\langle\bar{q}g_{s}\sigma
Gq\rangle(Q)\left[\frac{\alpha_{s}(Q)}{\alpha_{s}(\mu)}\right]^{\frac{2}{27}}\, ,\nonumber\\
\langle\bar{s}g_{s}\sigma
Gs\rangle(\mu)&=&\langle\bar{s}g_{s}\sigma
Gs\rangle(Q)\left[\frac{\alpha_{s}(Q)}{\alpha_{s}(\mu)}\right]^{\frac{2}{27}}\, ,\nonumber\\
m_{s}(\mu)&=&m_{s}\left(2\,\text{GeV}\right)\left[\frac{\alpha_{s}(\mu)}{\alpha_{s}(2\,\text{GeV})}\right]^{\frac{4}{9}}\, ,\nonumber\\
m_{c}(\mu)&=&m_{c}\left(m_{c}\right)\left[\frac{\alpha_{s}(\mu)}{\alpha_{s}(m_{c})}\right]^{\frac{12}{25}}\, ,\nonumber\\
\alpha_{s}(\mu)&=&\frac{1}{b_{0}t}\left[1-\frac{b_{1}}{b_{0}^{2}}\frac{\log t}{t}+\frac{b_{1}^{2}\left(\log^{2}t-\log t-1\right)+b_{0}b_{2}}{b_{0}^{4}t^{2}}\right]\, ,
\end{eqnarray}
where $t=\log \frac{\mu^{2}}{\Lambda^{2}}$, $b_{0}=\frac{33-2n_{f}}{12\pi}$, $b_{1}=\frac{153-19n_{f}}{24\pi^{2}}$, $b_{2}=\frac{2857-\frac{5033}{9}n_{f}+\frac{325}{27}n_{f}^{2}}{128\pi^{3}}$,
$\Lambda=213\,\text{MeV}$, $296\,\text{MeV}$ and $339\,\text{MeV}$
for the flavors $n_{f}=5$, $4$ and $3$, respectively \cite{XYZ}.

For the hadron mass, it is independent of the energy scale because of its observability.
However, in calculations,
the perturbative corrections are neglected, the operators of the orders $\mathcal{O}_n( \alpha_s^{k})$
with $k>1$ or the dimensions $n>10$ are discarded,
and the higher dimensional vacuum condensates
are factorized into lower dimensional ones
therefore the energy-scale dependence of the higher dimensional vacuum condensates is modified.
In addition, the variation of the heavy mass $m_c$ depending on the energy scale
leads to change of integral range $4m_c^2-s_0$ of the variable $ds$.
So we have to consider the energy-scale dependence of the QCD sum rules.

The hidden-charm four-quark system $c\bar{c}q'\bar{q}$
could be described by a double-well potential
with two light quarks $q'\bar{q}$ lying in the two wells respectively.
In the heavy quark limit, the $c$ quark can be taken as a static well potential,
which binds the light quark $q'$ to form a diquark
in the color antitriplet channel or binds the light antiquark $\bar{q}$
to form a meson in the color singlet channel (or a meson-like state in the color octet channel).
Then the hidden-charm four-quark states are characterized by the effective heavy quark mass ${\mathbb{M}}_{c}$
and the virtuality $V=\sqrt{M_{X/Y/Z}^{2}-\left(2{\mathbb{M}}_{c}\right)^{2}}$.
The effective mass ${\mathbb{M}}_{c}$ has uncertainties,
the optimal value in the diquark-antidiquark system
is not necessary the ideal value in the meson-meson system.
It is natural to take the energy scale $\mu=V$.
In this article, we use the energy-scale formula:
\begin{eqnarray}\label{Escalar}
\mu&=&\sqrt{M_{X/Y/Z}^{2}-\left(2{\mathbb{M}}_{c}\right)^{2}}\ ,
\end{eqnarray}
with the updated value of the effective $c$-quark mass ${\mathbb{M}}_{c}=1.85\,\rm{GeV}$
in the meson-meson molecular system to determine the ideal energy scales of the QCD spectral densities \cite{Wang-CPC-4390}.
For a better understanding of the energy-scale dependence in Eq. \eqref{Escalar},
one can refer to Ref. \cite{Z(3872),axialvector,energy-scale},
where the authors study the energy-scale dependence of the QCD sum rules for the hidden-charm tetraquark states
and molecular states in detail,
and suggest the above energy-scale formula for the first time.
In our calculations, we observe that the values of the masses $M_{Z_{1,2}}$
decrease slightly with  increase of the energy scales $\mu$
from QCD sum rules in Eq. \eqref{mass},
while Eq. \eqref{Escalar} indicates
that the values of the masses  $M_{Z_{1,2}}$ increase when the energy scales $\mu$ increase.
Thus there exist optimal energy scales, which lead to  reasonable  masses $M_{Z_{1,2}}$.

In Eq. \eqref{mass}, there are two free parameters: the Borel Parameter $T^2$ and the continuum threshold value $s_0$.
The extracted hadron mass is a function of the Borel parameter $T^2$ and the continuum threshold value $s_0$.
To obtain a reliable mass sum rule analysis,
we impose two criteria on the hidden-charm molecules
to choose suitable working ranges for these two free parameters.
The first criterion is the pole dominance on the phenomenological side,
which require the pole contributions (PCs) to be about $(40-60)\%$.
The PC is defined as:
\begin{eqnarray}
\text{PC}&=&\frac{\int_{4m_{c}^{2}}^{s_{0}}ds\rho_{1,2}\left(s\right)\exp\left(-\frac{s}{T^{2}}\right)} {\int_{4m_{c}^{2}}^{\infty}ds\rho_{1,2}\left(s\right)\exp\left(-\frac{s}{T^{2}}\right)}\ .
\end{eqnarray}
The second criterion is the convergence of the operator product expansion. To judge the
convergence, we calculate the contributions of the vacuum condensates $D(n)$
in the operator product expansion with the formula:
\begin{eqnarray}
D(n)&=&\frac{\int_{4m_{c}^{2}}^{s_{0}}ds\rho^{n}_{1,2}(s)\exp\left(-\frac{s}{T^{2}}\right)}
{\int_{4m_{c}^{2}}^{s_{0}}ds\rho_{1,2}\left(s\right)\exp\left(-\frac{s}{T^{2}}\right)}\ ,
\end{eqnarray}
where the index $n$ denotes the dimension of the vacuum condensates.

To search for the continuum threshold value $s_0$ more accurately,
we take into account the mass gaps
between the ground states and the first radial excited states,
which are usually taken as $(0.4-0.6)\,\text{GeV}$ in the four-quark sector.
For examples, the $Z(4430)$ is tentatively
assigned to be the first radial excitation of the $Z_{c}(3900)$
according to the analogous decays,
$Z_{c}(3900)^{\pm} \longrightarrow  J/\psi \pi^{\pm}$, $Z(4430)^{\pm} \longrightarrow \psi'\pi^{\pm}$
and the mass differences
$M_{Z(4430)}-M_{Z_{c}(3900)}=576\,\text{MeV}$, $M_{\psi'}-M_{J/\psi}=589\,\text{MeV}$ \cite{mass-gap1};
the $X(3915)$ and $X(4500)$ are assigned
to be the ground state and the first radial excited state of the $cs\bar{c}\bar{s}$ four-quark states, respectively,
and their mass difference is $M_{X(4500)}-M_{X(3915)}=588\,\text{MeV}$ \cite{mass-gap2}.
The relation
\begin{eqnarray}\label{s0-relation}
\sqrt{s_{0}}&=&M_{X/Y/Z}+(0.4-0.6)\,\text{GeV}\ ,
\end{eqnarray}
serves as a constraint on the masses of the hidden-charm four-quark states.

\begin{figure}[htp]
\centering
\includegraphics[totalheight=5cm,width=7cm]{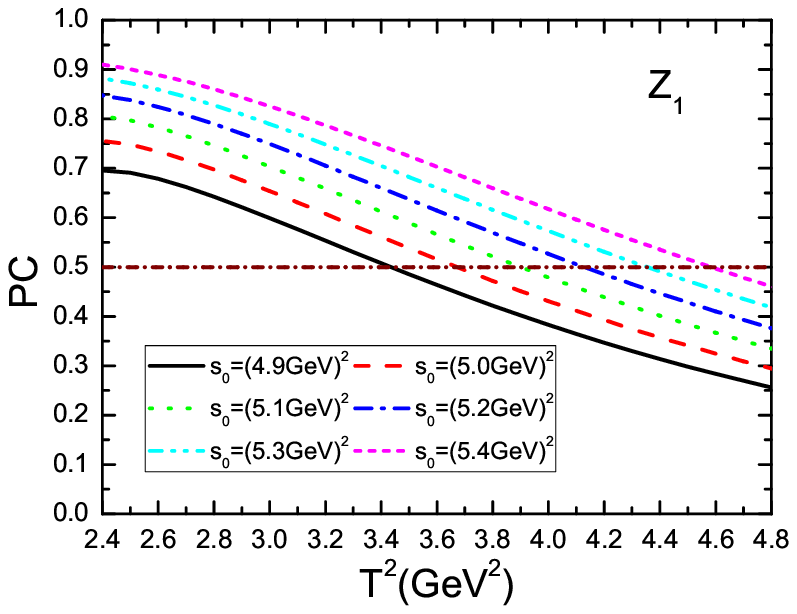}
\includegraphics[totalheight=5cm,width=7cm]{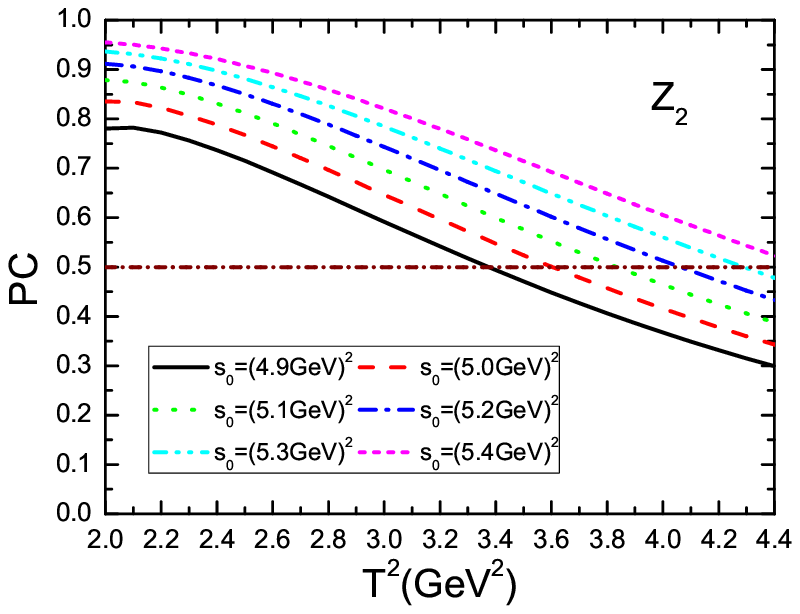}
\caption{The pole contributions with variations of the Borel parameter $T^{2}$ and the continuum threshold value $s_{0}$.}\label{fig:fig1}
\end{figure}

\begin{figure}[htp]
\centering
\includegraphics[totalheight=5cm,width=7cm]{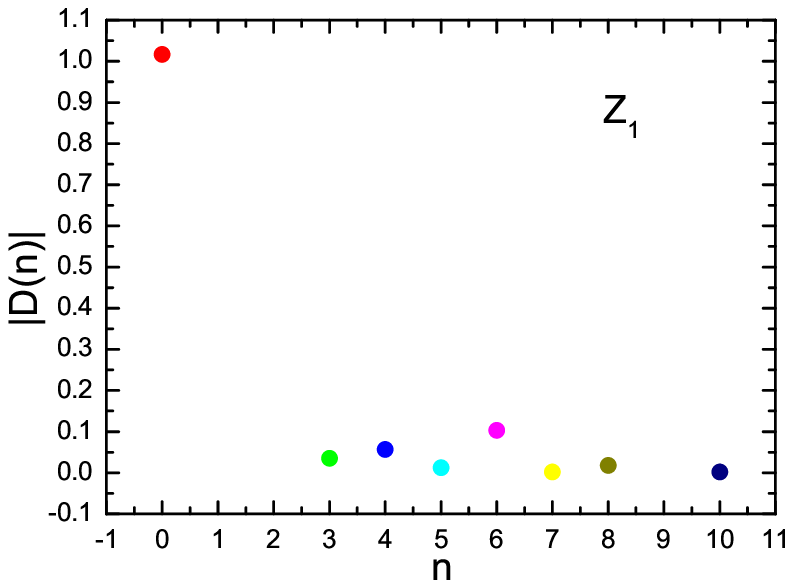}
\includegraphics[totalheight=5cm,width=7cm]{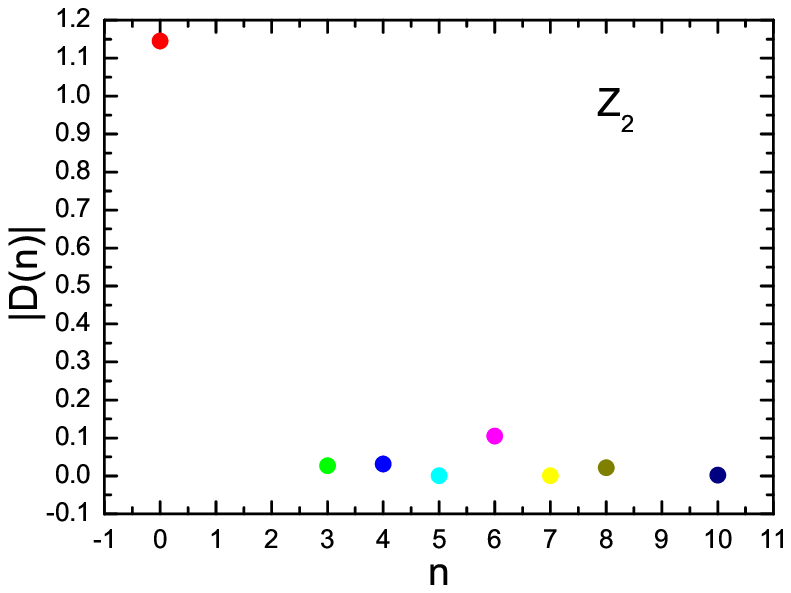}
\caption{The absolute contributions of the vacuum condensates with dimension $n$ in the operator product expansion.}\label{fig:fig2}
\end{figure}

\begin{figure}[htp]
\centering
\includegraphics[totalheight=5cm,width=7cm]{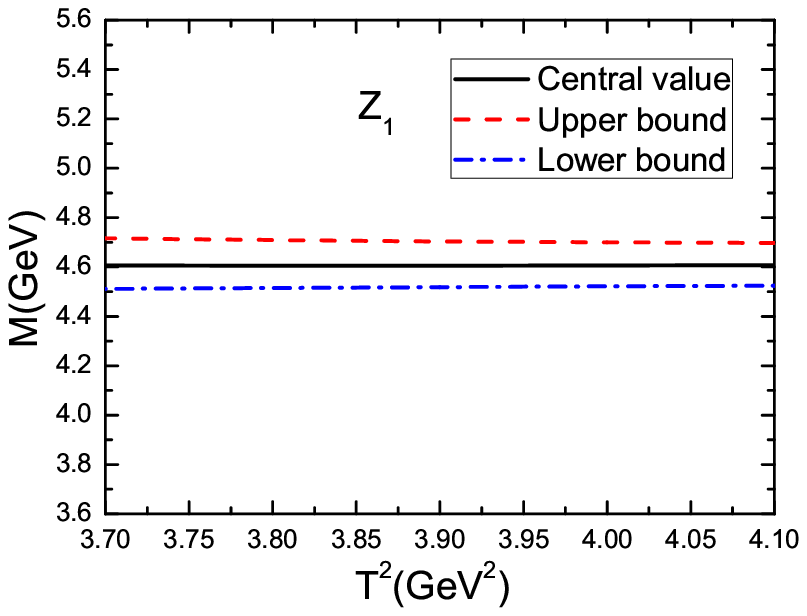}
\includegraphics[totalheight=5cm,width=7cm]{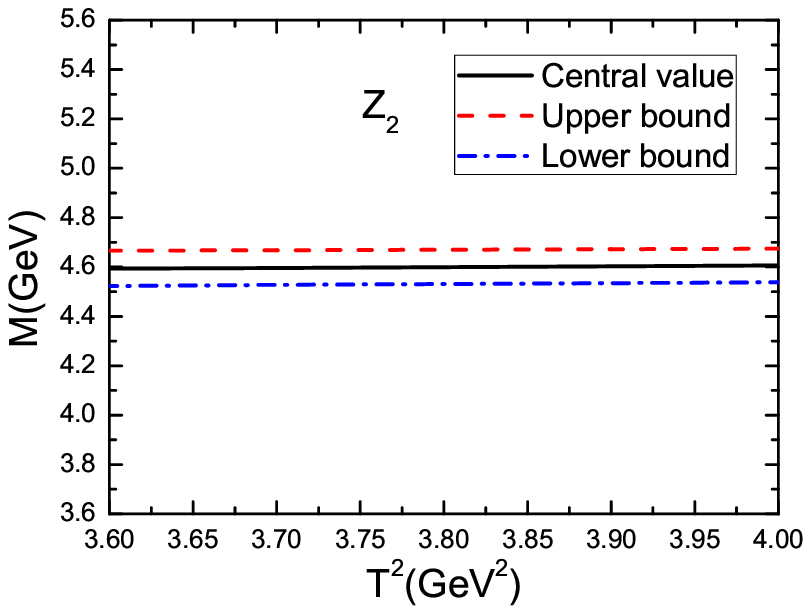}
\caption{The masses with variations of the Borel parameters $T^{2}$.}\label{fig:fig3}
\end{figure}

\begin{figure}[htp]
\centering
\includegraphics[totalheight=5cm,width=7cm]{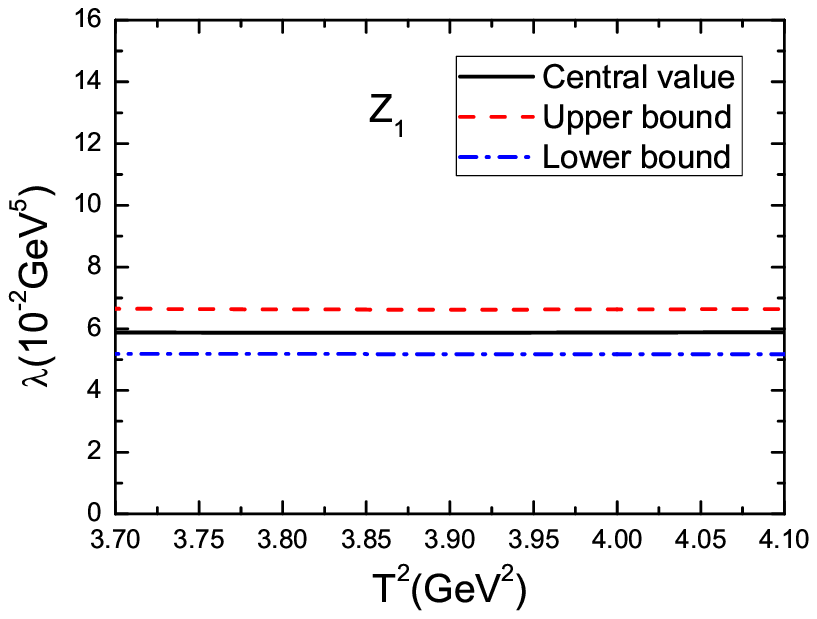}
\includegraphics[totalheight=5cm,width=7cm]{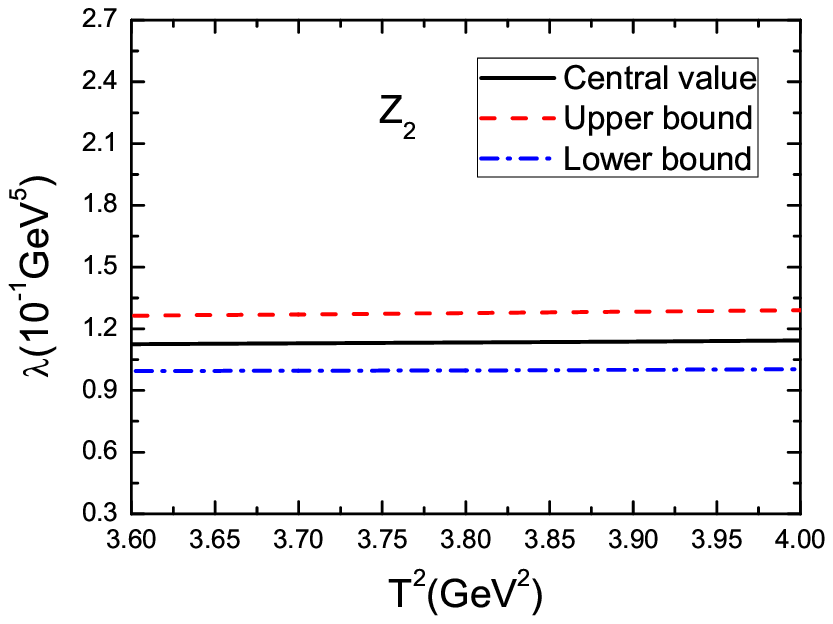}
\caption{The pole residues with variations of the Borel parameters $T^{2}$.}\label{fig:fig4}
\end{figure}

In Fig. \ref{fig:fig1},
we show the variations of the pole contributions with respect to the Borel parameters $T^2$
for different values of the continuum thresholds $s_0$
at the energy scales $\mu=2.7\,\rm{GeV}$ and $2.7\,\rm{GeV}$ for the states $Z_1$ and $Z_2$, respectively.
From the figure, we can see that the values $\sqrt{s_0}\leq4.9\,\rm{GeV}$
are too small to satisfy the pole dominance condition
and result in reasonable Borel windows for these two states $Z_{1,2}$.
To warrant the Borel platforms for the masses,
we take the values $T^2=(3.7-4.1)\,\rm{GeV}^2$ for the state $Z_1$
and $T^2=(3.6-4.0)\,\rm{GeV}^2$ for the state $Z_2$, respectively.
In the above Borel windows, if we choose the values $\sqrt{s_0}=(5.0-5.2)\,\rm{GeV}$,
the PCs are about $(42-60)\%$ and $(42-61)\%$ for the $Z_{1,2}$, respectively.
The pole dominance condition is well satisfied.

In Fig. \ref{fig:fig2},
we plot the absolute contributions of the vacuum condensates $|D(n)|$
in the operator product expansion at central values of the above input parameters for the states $Z_{1,2}$.
From the figure, we can observe that the dominant contributions come from the perturbative terms $D(0)$
for the $Z_{1,2}$.
The contributions of the other vacuum condensates are tiny.
The convergent behavior in the operator product expansion is very good.

\begin{table}[!ht]
\begin{center}
\begin{tabular}{|c|c|c|c|c|c|c|}
  \hline
   \hline
  & $\mu(\text{GeV})$ & $T^{2}(\text{GeV}^{2})$ & $\sqrt{s_{0}}(\text{GeV})$  & pole & $M_{Z}(\text{GeV})$ &  $\lambda_{Z}(\text{GeV}^{5})$\\
   \hline
   $Z_1: D\bar{D}_{s0}^*(2317)$ &$2.7$& $3.7-4.1$ & $5.1\pm0.1$ & $\left(42-60\right)\%$ & $4.61_{-0.08}^{+0.11}$ & $\left(5.87_{-0.69}^{+0.77}\right)\times10^{-2}$\\
   \hline
   $Z_2: D^*\bar{D}_{s1}^*(2460)$ &$2.7$& $3.6-4.0$ & $5.1\pm0.1$ & $\left(42-61\right)\%$ & $4.60_{-0.06}^{+0.07}$ & $\left(1.13_{-0.14}^{+0.16}\right)\times10^{-1}$\\
   \hline
  \hline
\end{tabular}
\end{center}
\caption{The energy scales, Borel parameters, continuum threshold parameters,
pole contributions, masses and pole residues for the molecular states.}\label{tab:tablenotes}
\end{table}

Thus, we obtain the Borel parameters, continuum thresholds and the pole contributions for
the states $Z_{1,2}$,
which are shown explicitly in Table \ref{tab:tablenotes}.
Taking into account all uncertainties of the input parameters,
we obtain the values of the masses and pole residues of the states $Z_{1,2}$,
which are shown in Table \ref{tab:tablenotes} and Figs. \ref{fig:fig3}--\ref{fig:fig4}.
From Table \ref{tab:tablenotes},
we cam see that the energy-scale formula $\mu =\sqrt{M_{X/Y/Z}^2-({2\mathbb{M}}_c)^2}$
and the relation $\sqrt{s_{0}}=M_{X/Y/Z}+(0.4-0.6)\,\text{GeV}$ are also well satisfied.

In the process of searching for the optimal Borel parameters and threshold parameters,
we went through a series of attempts and corrections.
Now we take a short digression to illustrate this procedure in more detail.
Taking the state $Z_1$ as an example,
firstly,
we set $M_{Z_1}=4.1\,\rm{GeV}$ tentatively
and obtain the energy scale $\mu=2.0\,\rm{GeV}$
according to the energy scale formula.
Then we take the continuum threshold parameter to be $\sqrt{s_0}=(4.2+0.5)\,\rm{GeV}$
as the energy gap between the ground state and the first radial excited state is about $(0.4-0.6)\,\rm{GeV}$,
and obtain the predicted masses $M_{Z_1}$, pole contributions
and the contributions of the vacuum condensates up to dimension-10.
We observe that the predicted masses $M_{Z_1}$ are much larger than $4.2\,\rm{GeV}$
and the pole contributions are much smaller than  $50\%$ in the region where the Borel platform appears,
furthermore, the contributions of the vacuum condensates of dimension 10 are not small enough.
Then we choose the mass $M_{Z_1}>4.2 \,\rm{GeV}$, meaning $M_{Z_1}=4.3 \,\rm{GeV}$, $4.4 \,\rm{GeV}$, $\cdots$
and reiterate the same procedure until obtain the ideal Borel parameters and continuum threshold parameters.
Finally, the predicted masses $M_{Z_{1,2}}$ and pole residues $\lambda_{Z_{1,2}}$  obtained by us are quite reliable.

The central value $M_{Z_1}=4.61\,\rm{GeV}$ is about $260\,\rm{MeV}$ above the threshold
$M_{D+\bar{D}_{s0}^*(2317)}=1870+2480=4350\,\rm{MeV}$,
where the mass of the $\bar{D}_{s0}^*(2317)$, $M_{\bar{D}_{s0}^*(2317)}=2480\,\rm{MeV}$,
is taken from the computed results of S. Godfrey and K. Moats
about excited charm and charm-strange mesons in Ref. \cite{Ds(2317)},
while the central value $M_{Z_2}=4.60\,\rm{GeV}$ is about $130\,\rm{MeV}$ above the threshold
$M_{D^*+\bar{D}_{s1}^*(2460)}=2010+2460=4470\,\rm{MeV}$.
The numerical results indicate that
the $D\bar{D}_{s0}^*(2317)$ and $D^*\bar{D}_{s1}^*(2460)$
are difficult to form bound state molecular states.

In Refs. \cite{Y(4274),DDs-molecules}, the authors study the analogous heavy meson systems.
In Ref. \cite{Y(4274)},  Liu, Luo and Zhu study the S-wave $D_s\bar{D}_{s0}^*(2317)$ system
through the heavy meson chiral perturbation theory,
considering the $\eta$ meson exchange between $D_s$ and $\bar{D}_{s0}^*(2317)$, which generates a potential to bind them,
and observe that there exists the $D_s\bar{D}_{s0}^*(2317)$  molecular state.
In  Ref. \cite{DDs-molecules}, similarly, using the heavy meson chiral perturbation theory,
Sanchez et al  study the S-wave $DD_{s0}^*(2317)$ and $D^*D_{s1}^*(2460)$ systems
exchanging a kaon to bind $D$($D^*$) and $D_{s0}^*$($D_{s1}^*$),
and predict the existence of $DD_{s0}^*(2317)$ and $D^*D_{s1}^*(2460)$  bound states.
Differently, in this article,
we construct the color singlet-singlet molecule-type interpolating currents $J_{1,2}(x)$
in Eqs. \eqref{eq1}--\eqref{eq2}
to study the $D\bar{D}_{s0}^*(2317)$ and $D^*\bar{D}_{s1}^*(2460)$ systems with QCD sum rules,
and give the prediction that
the $D\bar{D}_{s0}^*(2317)$ and $D^*\bar{D}_{s1}^*(2460)$
are difficult to form bound state molecular states.

Furthermore, the $Z_{1,2}$ are probably the resonance states,
since the constructed color singlet-singlet currents $J_{1,2}(x)$
may not necessarily correspond
the $D\bar{D}_{s0}^*(2317)$ and $D^*\bar{D}_{s1}^*(2460)$ bound state molecular states,
and could couple potentially to the $D\bar{D}_{s0}^*(2317)$ and $D^*\bar{D}_{s1}^*(2460)$ scattering states, respectively.
At the phenomenological side,
we can insert a complete set of intermediate hadronic states
with the same quantum numbers as the current operators $J_{1,2}(x)$
into the correlation functions $\Pi_{1,2}\left(p\right)$
to obtain the hadronic representations.
After isolating the $D\bar{D}_{s0}^*(2317)$ and $D^*\bar{D}_{s1}^*(2460)$ scattering states,
we get the following results,
\begin{eqnarray}
\Pi_{1}\left(p\right)
&=&\frac{i}{\left(2\pi\right)^4}\int d^4k \frac{i}{k^2-M_D^2}\frac{i}{\left(k+p\right)^2-M^2_{\bar{D}_{s0}^*}}
\left\{4\frac{f_D^2f^2_{\bar{D}_{s0}^*}k^4\left(k+p\right)^2}{\left(m_c+m_q\right)^2} \right. \nonumber\\
&&\left.+f_D^2f^2_{\bar{D}_{s0}^*}\left[k\cdot(k+p)\right]^2
+4\frac{f_D^2f^2_{\bar{D}_{s0}^*}k^2\sqrt{(k+p)^2}k\cdot(k+p)}{m_c+m_q}\right\}+\cdots\, ,\nonumber\\
\Pi_{2}\left(p\right)
&=&\frac{i}{\left(2\pi\right)^4}\int d^4k \frac{i}{k^2-M_{D^*}^2}\frac{i}{\left(k+p\right)^2-M^2_{\bar{D}_{s1}^*}}f_{D^*}^2f^2_{\bar{D}_{s1}^*}k^2\left(k+p\right)^2
\left[g_{\mu \nu}-\frac{k_\mu k_\nu}{k^2}\right] \nonumber\\
&&\times\left[g^{\mu \nu}-\frac{(k+p)^\mu (k+p)^\nu}{(k+p)^2}\right]+\cdots\, ,
\end{eqnarray}
where the decay constants$f_D$, $f_{D^*}$, $f_{\bar{D}_{s0}^*}$ and $f_{\bar{D}_{s1}^*}$ are defined by
\begin{eqnarray}
\langle0|J_{1}(0)|D(k)\bar{D}_{s0}^*(k+p)\rangle &=& 2\frac{f_D k^2}{m_c+m_q}
f_{\bar{D}_{s0}^*}\sqrt{(k+p)^2}+f_D f_{\bar{D}_{s0}^*}k\cdot(k+p)\, , \nonumber\\
\langle0|J_{2}(0)|D^*(k)\bar{D}_{s0}^*(k+p)\rangle &=& f_{D^*} f_{\bar{D}_{s1}^*}\sqrt{k^2(k+p)^2}\varepsilon_\mu \varepsilon^\mu \, ,
\end{eqnarray}
the $\varepsilon_\mu$ are the polarization vectors of the $D^*$ and $\bar{D}_{s1}^*$.

\begin{figure}[htp]
\centering
\includegraphics[totalheight=5cm,width=7cm]{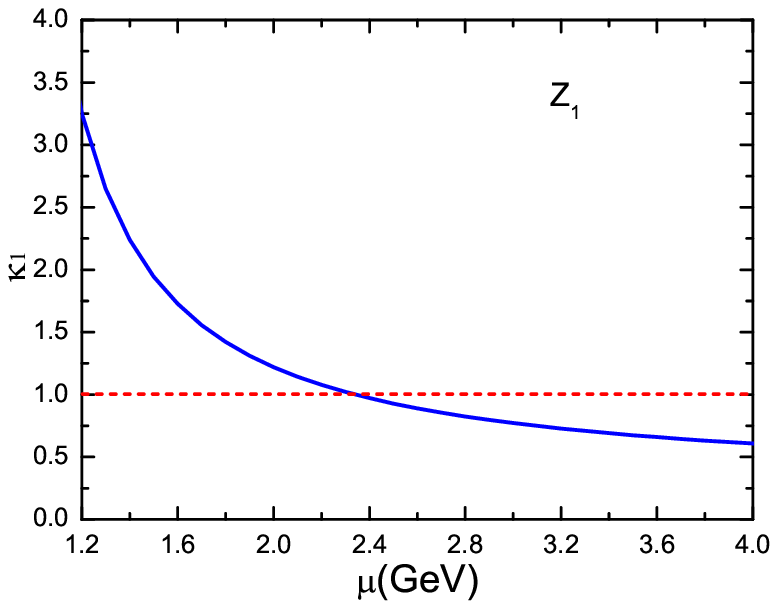}
\includegraphics[totalheight=5cm,width=7cm]{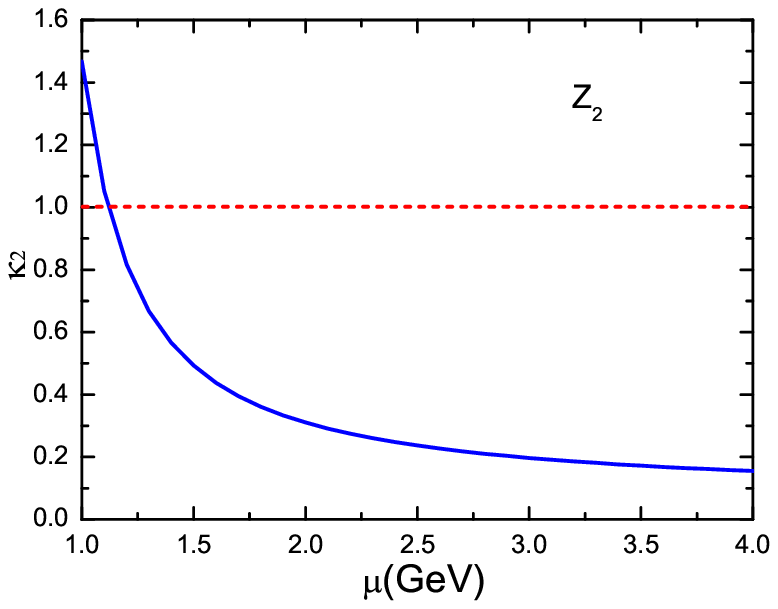}
\caption{The coefficient $\kappa_i$ with variation of the energy scale $\mu$.}\label{fig:fig5}
\end{figure}

We rewrite the correlation functions $\Pi_{1,2}(p)$ into the following forms through dispersion relation,
\begin{eqnarray}
\Pi_{1}\left(p\right)
&=&\frac{f_D^2f^2_{\bar{D}_{s0}^*}}{16\pi^2}\int^{s_0}_{\left(M_D+M_{\bar{D}_{s0}^*}\right)^2}ds\frac{1}{s-p^2}
\sqrt{\left(\frac{M^2_{\bar{D}_{s0}^*}-M_D^2-s}{s}\right)^2-\frac{4M_D^2}{s}} \nonumber\\
&&\times\left[\frac{4M_D^4M^2_{\bar{D}_{s0}^*}}{\left(m_c+m_q\right)^2}
+2M_D^2M_{\bar{D}_{s0}^*}\left(\frac{M_D^2+M^2_{\bar{D}_{s0}^*}-s}{m_c+m_q}\right)
+\left(\frac{M_D^2+M^2_{\bar{D}_{s0}^*}-s}{2}\right)^2\right] \nonumber\\
&&+\cdots\, , \nonumber\\
\Pi_{2}\left(p\right)
&=&\frac{f_{D^*}^2f^2_{\bar{D}_{s1}^*}}{16\pi^2}\int^{s_0}_{\left(M_{D^*}+M_{\bar{D}_{s1}^*}\right)^2}ds\frac{1}{s-p^2}
\sqrt{\left(\frac{M^2_{\bar{D}_{s1}^*}-M_{D^*}^2-s}{s}\right)^2-\frac{4M_{D^*}^2}{s}} \nonumber\\
&&\times M_{D^*}^2M^2_{\bar{D}_{s1}^*}
\left[2+\frac{\left(M_{D^*}^2+M^2_{\bar{D}_{s1}^*}-s\right)^2}{4M_{D^*}^2M^2_{\bar{D}_{s1}^*}}\right]+\cdots\, .
\end{eqnarray}
In this article,
we choose the value $s_0 >\left(M_D+M_{\bar{D}_{s0}^*}\right)^2$, $\left(M_{D^*}+M_{\bar{D}_{s1}^*}\right)^2$,
the QCD sum rules can be written as
\begin{eqnarray}
&&\frac{f_D^2f^2_{\bar{D}_{s0}^*}}{16\pi^2}\int^{s_0}_{\left(M_D+M_{\bar{D}_{s0}^*}\right)^2}ds
\sqrt{\left(\frac{M^2_{\bar{D}_{s0}^*}-M_D^2-s}{s}\right)^2-\frac{4M_D^2}{s}} \nonumber\\
&&\times\left[\frac{4M_D^4M^2_{\bar{D}_{s0}^*}}{\left(m_c+m_q\right)^2}
+2M_D^2M_{\bar{D}_{s0}^*}\left(\frac{M_D^2+M^2_{\bar{D}_{s0}^*}-s}{m_c+m_q}\right)
+\left(\frac{M_D^2+M^2_{\bar{D}_{s0}^*}-s}{2}\right)^2\right] \exp\left(-\frac{s}{T^2}\right)\nonumber\\
&&=\kappa_1 \,\int_{4m_c^2}^{s_0} ds  \rho_1(s)  \,\exp\left(-\frac{s}{T^2}\right) \, , \nonumber\\
&&\frac{f_{D^*}^2f^2_{\bar{D}_{s1}^*}}{16\pi^2}\int^{s_0}_{\left(M_{D^*}+M_{\bar{D}_{s1}^*}\right)^2}ds
\sqrt{\left(\frac{M^2_{\bar{D}_{s1}^*}-M_{D^*}^2-s}{s}\right)^2-\frac{4M_{D^*}^2}{s}} \nonumber\\
&&\times M_{D^*}^2M^2_{\bar{D}_{s1}^*}
\left[2+\frac{\left(M_{D^*}^2+M^2_{\bar{D}_{s1}^*}-s\right)^2}{4M_{D^*}^2M^2_{\bar{D}_{s1}^*}}\right]\exp\left(-\frac{s}{T^2}\right) \nonumber\\
&&=\kappa_2 \,\int_{4m_c^2}^{s_0} ds  \rho_2(s)  \,\exp\left(-\frac{s}{T^2}\right) \,,
\end{eqnarray}
where we introduce a coefficient $\kappa_i$, if $\kappa_i =1$,
the QCD sum rules can be saturated by the scattering states
$D\bar{D}_{s0}^*(2317)$ and $D^*\bar{D}_{s1}^*(2460)$, respectively.
The input parameters are taken as $M_D=1.87\,\text{GeV}$, $M_{D^*}=2.01\,\text{GeV}$, $M_{\bar{D}^*_{s1}}=2.46\,\text{GeV}$ \cite{XYZ},
$M_{\bar{D}^*_{s0}}=2.48\,\text{GeV}$ \cite{Ds(2317)},
$f_D =0.208 \,\text{GeV}$, $f_{D^*}=0.263\,\text{GeV}$,
$f_{\bar{D}^*_{s0}}=0.333\,\text{GeV}$, $f_{\bar{D}^*_{s1}}=0.245\,\text{GeV}$ \cite{EPJC75(427)},
$s_0 = 5.1^2 \,\text{GeV}^2$.
In Fig. \ref{fig:fig5},
we plot the coefficient $\kappa_i$ with variation of the energy scale $\mu$
at $T^2=3.9\,\text{GeV}^2$ and $3.8\,\text{GeV}^2$ for the $Z_{1,2}$, respectively.
At the vicinities of the energy scale $\mu=2.3\,\text{GeV}$ and $1.1\,\text{GeV}$,
$\kappa_i\approx 1$,
however,
from the figure, we can see that
the coefficient $\kappa_i$ decreases monotonously with increase of the energy scale $\mu$.
The reliable QCD sum rules do not depend heavily on the energy scale $\mu$.
So, the QCD sum rules can not  be saturated by the scattering states
$D\bar{D}_{s0}^*(2317)$ and $D^*\bar{D}_{s1}^*(2460)$, respectively.

In the following, we perform Fierz re-arrangement for the currents $J_{1,2}$
both in the color space and Dirac-spinor space to obtain the results,
\begin{eqnarray}
J_1 &=& -\frac{1}{12}\bar{c}i\gamma_5 c\bar{q}s
-\frac{i}{12}\bar{c}\gamma_\beta \gamma_5 c\bar{q}\gamma^\beta s
-\frac{i}{24}\bar{c}\sigma_{\beta \tau} \gamma_5 c\bar{q}\sigma^{\beta \tau}s
+\frac{i}{12}\bar{c}\gamma_\beta c\bar{q}\gamma^\beta \gamma_5 s
-\frac{1}{12}\bar{c}c\bar{q}i\gamma_5 s   \nonumber\\
&&-\frac{1}{8}\bar{c}i\gamma_5 \lambda^a c\bar{q}\lambda^a s
-\frac{i}{8}\bar{c}\gamma_\beta \gamma_5 \lambda^a c\bar{q}\gamma^\beta \lambda^a s
-\frac{i}{16}\bar{c}\sigma_{\beta \tau} \gamma_5 \lambda^a c\bar{q}\sigma^{\beta \tau}\lambda^a s
+\frac{i}{8}\bar{c}\gamma_\beta \lambda^a c\bar{q}\gamma^\beta \gamma_5 \lambda^a s  \nonumber\\
&&-\frac{1}{8}\bar{c}\lambda^a c\bar{q}i\gamma_5 \lambda^a s \, , \nonumber\\
J_2 &=&-\frac{i}{3}\bar{c}i\gamma_5 c\bar{q}s
+\frac{1}{6}\bar{c}\gamma_\beta \gamma_5 c\bar{q}\gamma^\beta s
+\frac{1}{6}\bar{c}\gamma_\beta  c\bar{q}\gamma^\beta \gamma_5s
+\frac{i}{3}\bar{c} c\bar{q}i\gamma_5s   \nonumber\\
&&-\frac{i}{2}\bar{c}i\gamma_5 \lambda^a c\bar{q}\lambda^a s
+\frac{1}{4}\bar{c}\gamma_\beta \gamma_5 \lambda^a c\bar{q}\gamma^\beta \lambda^a s
+\frac{1}{4}\bar{c}\gamma_\beta  \lambda^a c\bar{q}\gamma^\beta \gamma_5\lambda^a s
+\frac{i}{2}\bar{c} \lambda^a c\bar{q}i\gamma_5\lambda^a s \,.
\end{eqnarray}
The components $\bar{c}\Gamma c\bar{q}\Gamma's$ and $\bar{c}\Gamma \lambda^a c\bar{q}\Gamma'\lambda^a s$
couple potentially to a series of charmonium-light-meson pairs or charmonium-like molecular states
or charmonium-like molecule-like states,
where $\Gamma$, $\Gamma'=1$, $\gamma_\beta$, $\gamma_\beta \gamma_5$, $i\gamma_5$,
$\sigma_{\beta \tau}$, $\sigma_{\beta \tau}\gamma_5$.
For example,
the current $J_1$ couples potentially to the meson pairs through its components,
\begin{eqnarray}
\bar{c}i\gamma_5 c\bar{q}s & \propto & \eta_c K^*_0\,,  \cdots \, , \nonumber\\
\bar{c}c\bar{q}i\gamma_5 s & \propto & \chi_{c0}K \,,\cdots \, , \nonumber\\
\bar{c}\gamma_\beta \gamma_5 c\bar{q}\gamma^\beta s & \propto & \chi_{c1}K^*\,, h_{c}K^*\,,\cdots \, , \nonumber\\
\bar{c}\gamma_\beta c\bar{q}\gamma^\beta \gamma_5 s  & \propto & J/\psi K_1\,, \cdots \, , \nonumber\\
\bar{c}\sigma_{\beta \tau} \gamma_5 c\bar{q}\sigma^{\beta \tau}s  & \propto &  J/\psi K_1\,,h_{c}K^*\,, \cdots \, .
\end{eqnarray}
We cannot distinguish those contributions to study them exclusively.
Hence,
we infer that the  $Z_{1,2}$ are particular resonance states,
which are the special superpositions of the scattering states, molecular states and molecule-like states,
and embody the net effects.
Moreover, for the meson-meson scattering states such as $\eta_c K^*_0$, $\chi_{c0}K$, $J/\psi K_1$, $\cdots $
lying below the $Z_{1}$,
the $Z_{1}$ can decay to them easily through fall-apart mechanism,
and the decays contribute a finite width to the $Z_{1}$.
Now, we discuss an effect of the finite width on the predicted mass $M_{Z_{1}}$.
We consider the contributions of the meson-loops to the correlation function $\Pi_1(p)$,
as the current $J_1(x)$ has non-vanishing couplings with the scattering states
$\eta_c K^*_0$, $\chi_{c0}K$, $J/\psi K_1$, etc.
\begin{eqnarray}
\Pi_{1}\left(p\right)&=&-\frac{\widehat{\lambda}_{Z_1}^2}{p^2
-\widehat{M}_{Z_1}^2-\Sigma_{\eta_c K^*_0}(p)-\Sigma_{\chi_{c0}K}(p)-\Sigma_{J/\psi K_1}(p)+\cdots}+\cdots\, ,
\end{eqnarray}
where the $\widehat{\lambda}_{Z_1}$ and $\widehat{M}_{Z_1}$
are bare quantities to absorb the divergences
in the self-energies $\Sigma_{\eta_c K^*_0}(p)$, $\Sigma_{\chi_{c0}K}(p)$, $\Sigma_{J/\psi K_1}(p)$, etc.
The renormalized self-energies contribute a finite imaginary part to modify the dispersion relation,
\begin{eqnarray}
\Pi_{1}\left(p\right)&=&-\frac{\lambda_{Z_1}^2}{p^2
-M_{Z_1}^2+i\sqrt{p^2}\Gamma(p^2)}+\cdots\, .
\end{eqnarray}
The finite width effect is considered through the following simple change
in the hadronic spectral density,
\begin{eqnarray}
\delta\left(s-M_{Z_1}^2\right)&\rightarrow&
\frac{1}{\pi}\frac{\sqrt{s}\Gamma_{Z_1}\left(s\right)}{\left(s-M^2_{Z_1}\right)^2+s\Gamma^2_{Z_1}\left(s\right)}\, ,
\end{eqnarray}
where
\begin{eqnarray}
\Gamma_{Z_1}\left(s\right)&=&
\Gamma_{Z_1}\frac{M^2_{Z_1}}{s}\, .
\end{eqnarray}
We take the central values of the input parameters, and $\Gamma_{Z_1}=300\,\text{MeV}$(not small).
Then the phenomenological side of the QCD sum rules in Eq. \eqref{PoleResidue}
changes as follows,
\begin{eqnarray}
B_{T^2}\Pi_1&=&
\lambda^2_{Z_1}\exp\left(-\frac{M^2_{Z_1}}{T^2}\right) \nonumber\\
&\rightarrow&\frac{\lambda^2_{Z_1}}{\pi}\int^{s_0}_{(M_{J/\psi}+M_{K_1})^2} ds\frac{\sqrt{s}\Gamma_{Z_1}\left(s\right)}{\left(s-M^2_{Z_1}\right)^2+s\Gamma^2_{Z_1}\left(s\right)}
\exp\left(-\frac{s}{T^2}\right) \nonumber\\
&=&0.70\lambda^2_{Z_1}\exp\left(-\frac{M^2_{Z_1}}{T^2}\right)
\end{eqnarray}
and
\begin{eqnarray}
-\frac{1}{d\left(1/T^2\right)}B_{T^2}\Pi_1&=&
M^2_{Z_1}\lambda^2_{Z_1}\exp\left(-\frac{M^2_{Z_1}}{T^2}\right) \nonumber\\
&\rightarrow&\frac{\lambda^2_{Z_1}}{\pi}\int^{s_0}_{(M_{J/\psi}+M_{K_1})^2} ds \, s\frac{\sqrt{s}\Gamma_{Z_1}\left(s\right)}{\left(s-M^2_{Z_1}\right)^2+s\Gamma^2_{Z_1}\left(s\right)}
\exp\left(-\frac{s}{T^2}\right) \nonumber\\
&=&0.70M^2_{Z_1}\lambda^2_{Z_1}\exp\left(-\frac{M^2_{Z_1}}{T^2}\right)\, ,
\end{eqnarray}
where the $B_{T^2}$ denotes the Borel transformation.
The numerical factor 0.70 can be absorbed safely into the pole residue $\lambda_{Z_1}$.
Therefore, in this article,
when we take the zero width approximation in Eq. \eqref{correlation-function-1},
the predicted masses $M_{Z_{1,2}}$ are reasonable.

\section{Conclusion}
In this article, we assume that
there exist the pseudoscalar $D\bar{D}_{s0}^*(2317)$ and $D^*\bar{D}_{s1}^*(2460)$ molecular states $Z_{1,2}$,
and study their masses with
the color singlet-singlet interpolating currents through the QCD sum rule approach.
In calculations, we carry out
the operator product expansion up to the vacuum condensates of dimension 10
and use the formula
$\mu=\sqrt{M_{X/Y/Z}^{2}-\left(2{\mathbb{M}}_{c}\right)^{2}}$
to determine the energy scales of the QCD spectral densities.
The numerical results show that the central value of the state $Z_1$,
$M_{Z_1}=4.61\,\rm{GeV}$,
is about $260\,\rm{MeV}$ above the $D\bar{D}_{s0}^*(2317)$ threshold,
while, in the case of the $Z_2$, the central value
$M_{Z_2}=4.61\,\rm{GeV}$ is about $130\,\rm{MeV}$ above the $D^*\bar{D}_{s1}^*(2460)$ threshold,
which indicate that the $D\bar{D}_{s0}^*(2317)$ and $D^*\bar{D}_{s1}^*(2460)$
are difficult to form   bound state  molecular states.
The  $Z_{1,2}$ are probably particular resonance states,
which are the special superpositions of the scattering states, molecular states and molecule-like states,
and embody the net effects.
We expect that these results in our work could be helpful for investigating the $Z_{1,2}$ experimentally,
and would be able to be testified in the future experiments, such as BESIII, LHCb and Belle-II.

\section*{Acknowledgements}
This work is supported by National Natural Science Foundation, Grant Number 11775079.

\section*{Appendix}
The explicit expressions of the QCD spectral densities $\rho_{1,2}\left(s\right)$,
\begin{eqnarray}
\rho^{0}_{1}(s)&=&\frac{3}{2048\pi^{6}}\int_{y_{i}}^{y_{f}}dy\int_{z_{i}}^{1-y}dz\ yz\left(1-y-z\right)^{3}\left(s-\hat{m}_{c}^{2}\right)^{2}\left(7s^2-6s\hat{m}_{c}^{2}+\hat{m}_{c}^{4}\right) \nonumber\\
&&-\frac{3m_s m_{c}}{2048\pi^{6}}\int_{y_{i}}^{y_{f}}dy\int_{z_{i}}^{1-y}dz\, \left(y+z\right)\left(1-y-z\right)^{2}\left(s-\hat{m}_{c}^{2}\right)^{2}\left(5s-2\hat{m}_{c}^{2}\right)\ ,
\end{eqnarray}

\begin{eqnarray}
\rho^{0}_{2}(s)&=&\frac{3}{1024\pi^{6}}\int_{y_{i}}^{y_{f}}dy\int_{z_{i}}^{1-y}dz\ yz\left(1-y-z\right)^{3}\left(s-\hat{m}_{c}^{2}\right)^{2}\left(7s^2-6s\hat{m}_{c}^{2}+\hat{m}_{c}^{4}\right) \nonumber\\
&&+\frac{3}{1024\pi^{6}}\int_{y_{i}}^{y_{f}}dy\int_{z_{i}}^{1-y}dz\
yz\left(1-y-z\right)^{2}\left(s-\hat{m}_{c}^{2}\right)^3\left(3s-\hat{m}_{c}^{2}\right) \nonumber\\
&&-\frac{3m_s m_c}{1024\pi^{6}}\int_{y_{i}}^{y_{f}}dy\int_{z_{i}}^{1-y}dz
\left(y+z\right)\left(1-y-z\right)^{2}\left(s-\hat{m}_{c}^{2}\right)^2\left(5s-2\hat{m}_{c}^{2}\right)\ ,
\end{eqnarray}

\begin{eqnarray}
\rho^{3}_{1}(s)&=&\frac{3m_{c}\left(\langle\bar{s}s\rangle-\langle\bar{q}q\rangle\right)}{128\pi^{4}}\int_{y_{i}}^{y_{f}}dy\int_{z_{i}}^{1-y}dz\, \left(y+z\right)\left(1-y-z\right)\left(s-\hat{m}_{c}^{2}\right)\left(2s-\hat{m}_{c}^{2}\right) \nonumber\\
&&+\frac{3m_s\langle\bar{s}s\rangle}{128\pi^4}\int_{y_{i}}^{y_{f}}dy\int_{z_{i}}^{1-y}dz\,
yz\left(1-y-z\right)\left(10s^2-12s\hat{m}_{c}^{2}+3\hat{m}_{c}^{4}\right) \nonumber\\
&&+\frac{3m_s m_c^2\langle\bar{q}q\rangle}{64\pi^4}\int_{y_{i}}^{y_{f}}dy\int_{z_{i}}^{1-y}dz\, \left(s-\hat{m}_{c}^{2}\right)\ ,
\end{eqnarray}

\begin{eqnarray}
\rho^{3}_{2}(s)&=&\frac{3m_{c}\left(\langle\bar{s}s\rangle-\langle\bar{q}q\rangle\right)}{64\pi^{4}}
\int_{y_{i}}^{y_{f}}dy\int_{z_{i}}^{1-y}dz \left(y+z\right)\left(1-y-z\right)\left(s-\hat{m}_{c}^{2}\right)
\left(2s-\hat{m}_{c}^{2}\right) \nonumber\\
&&+\frac{3m_s \langle\bar{s}s\rangle}{64\pi^{4}}\int_{y_{i}}^{y_{f}}dy\int_{z_{i}}^{1-y}dz\,
yz\left(1-y-z\right)\left(10s^2-12s\hat{m}_c^2+3\hat{m}_c^4\right) \nonumber\\
&&+\frac{3m_s \langle\bar{s}s\rangle}{64\pi^{4}}\int_{y_{i}}^{y_{f}}dy\int_{z_{i}}^{1-y}dz\,
yz\left(s-\hat{m}_{c}^{2}\right)\left(2s-\hat{m}_{c}^{2}\right) \nonumber\\
&&+\frac{3m_s m_c^2 \langle\bar{q}q\rangle}{16\pi^{4}}\int_{y_{i}}^{y_{f}}dy\int_{z_{i}}^{1-y}dz\,
\left(s-\hat{m}_{c}^{2}\right)\,
\end{eqnarray}

\begin{eqnarray}
\rho^{4}_{1}(s)&=&\frac{m_s m_c^3}{2048\pi^{4}}\langle\frac{\alpha_{s}GG}{\pi}\rangle\int_{y_{i}}^{y_{f}}dy\int_{z_{i}}^{1-y}dz\ \left(\frac{y}{z^3}+\frac{z}{y^3}+\frac{1}{y^2}+\frac{1}{z^2}\right)
\left(1-y-z\right)^{2}\left[2+s\delta(s-\hat{m}_{c}^{2})\right] \nonumber\\
&&-\frac{m_{c}^{2}}{512\pi^{4}}\langle\frac{\alpha_{s}GG}{\pi}\rangle\int_{y_{i}}^{y_{f}}dy\int_{z_{i}}^{1-y}dz\ \left(\frac{z}{y^{2}}+\frac{y}{z^{2}}\right)\left(1-y-z\right)^{3}
\left[2s-\hat{m}_{c}^{2}+\frac{s^2}{6}\delta(s-\hat{m}_{c}^{2})\right]\nonumber\\
&&-\frac{3m_s m_{c}}{2048\pi^{4}}\langle\frac{\alpha_{s}GG}{\pi}\rangle\int_{y_{i}}^{y_{f}}dy\int_{z_{i}}^{1-y}dz\
\left[\left(\frac{y}{z^2}+\frac{z}{y^2}\right)\left(1-y-z\right)+4\right]
\left(1-y-z\right)\left(3s-\hat{m}_{c}^{2}\right)\nonumber\\
&&+\frac{3}{1024\pi^{4}}\langle\frac{\alpha_{s}GG}{\pi}\rangle\int_{y_{i}}^{y_{f}}dy\int_{z_{i}}^{1-y}dz\ \left(y+z\right)\left(1-y-z\right)^{2}\left(10s^2-12s\hat{m}_{c}^{2}+3\hat{m}_{c}^{4}\right) \ ,
\end{eqnarray}

\begin{eqnarray}
\rho^{4}_{2}(s)&=&\frac{m_s m_{c}^{3}}{1024\pi^{4}}\langle\frac{\alpha_{s}GG}{\pi}\rangle
\int_{y_{i}}^{y_{f}}dy\int_{z_{i}}^{1-y}dz
\left(\frac{y}{z^3}+\frac{z}{y^3}+\frac{1}{y^2}+\frac{1}{z^2}\right)\left(1-y-z\right)^2
\left[2+s\delta\left(s-\hat{m}_{c}^{2}\right)\right]\nonumber\\
&&-\frac{\left(2m_{c}^{2}+3m_s m_c\right)}{1024\pi^{4}}\langle\frac{\alpha_{s}GG}{\pi}\rangle
\int_{y_{i}}^{y_{f}}dy\int_{z_{i}}^{1-y}dz\ \left(\frac{y}{z^{2}}+\frac{z}{y^{2}}\right)\left(1-y-z\right)^{2}
\left(3s-2\hat{m}_{c}^{2}\right) \nonumber\\
&&-\frac{m_c^2}{256\pi^{4}}\langle\frac{\alpha_{s}GG}{\pi}\rangle
\int_{y_{i}}^{y_{f}}dy\int_{z_{i}}^{1-y}dz\ \left(\frac{y}{z^{2}}+\frac{z}{y^{2}}\right)\left(1-y-z\right)^{3}
\left[2s-\tilde{m}_c^2+\frac{s^2}{6}\delta\left(s-\hat{m}_{c}^{2}\right)\right]  \nonumber\\
&&+\frac{1}{256\pi^{4}}\langle\frac{\alpha_{s}GG}{\pi}\rangle\int_{y_{i}}^{y_{f}}dy\int_{z_{i}}^{1-y}dz\ \left(y+z\right)\left(1-y-z\right)\left(s-\hat{m}_{c}^{2}\right)\left(2s-\hat{m}_{c}^{2}\right) \nonumber\\
&&-\frac{1}{512\pi^{4}}\langle\frac{\alpha_{s}GG}{\pi}\rangle\int_{y_{i}}^{y_{f}}dy\int_{z_{i}}^{1-y}dz\ \left(y+z\right)\left(1-y-z\right)^2\left(10s^2-12s\hat{m}_{c}^{2}+3\hat{m}_{c}^{4}\right) \nonumber\\
&&+\frac{3m_s m_{c}}{256\pi^{4}}\langle\frac{\alpha_{s}GG}{\pi}\rangle\int_{y_{i}}^{y_{f}}dy\int_{z_{i}}^{1-y}dz
\left(1-y-z\right)
\left(3s-2\hat{m}_{c}^{2}\right)\ ,
\end{eqnarray}

\begin{eqnarray}
\rho^{5}_{1}(s)&=&\frac{3m_{c}\left(\langle\bar{q}g_{s}\sigma Gq\rangle-\langle\bar{s}g_{s}\sigma Gs\rangle\right)}{512\pi^{4}}\int_{y_{i}}^{y_{f}}dy\int_{z_{i}}^{1-y}dz\
\left[\left(y+z\right)-2\left(\frac{y}{z}+\frac{z}{y}\right)\left(1-y-z\right)\right]\nonumber\\
&&\left(3s-2\hat{m}_{c}^{2}\right)
-\frac{3m_s\langle\bar{s}g_{s}\sigma Gs\rangle}{128\pi^{4}}\int_{y_{i}}^{y_{f}}dy\int_{z_{i}}^{1-y}dz\
yz\left[2s-\hat{m}_{c}^{2}+\frac{s^2}{6}\delta(s-\hat{m}_{c}^{2})\right] \nonumber\\
&&+\frac{3m_{s}\langle\bar{q}g_{s}\sigma Gq\rangle}{256\pi^{4}}\int_{y_{i}}^{y_{f}}dy\int_{z_{i}}^{1-y}dz\
\hat{m}_{c}^{2}-\frac{3m_s m_c^2\langle\bar{q}g_{s}\sigma Gq\rangle}{256\pi^{4}}\int_{y_{i}}^{y_{f}}dy\ ,
\end{eqnarray}

\begin{eqnarray}
\rho^{5}_{2}(s)&=&\frac{3m_{c}\left(\langle\bar{q}g_{s}\sigma Gq\rangle-\langle\bar{s}g_{s}\sigma Gs\rangle\right)}{256\pi^{4}}\int_{y_{i}}^{y_{f}}dy\int_{z_{i}}^{1-y}dz\ \left(y+z\right)
\left(3s-2\hat{m}_{c}^{2}\right) \nonumber\\
&&-\frac{3m_s\langle\bar{s}g_{s}\sigma Gs\rangle}{64\pi^{4}} \int_{y_{i}}^{y_{f}}dy\int_{z_{i}}^{1-y}dz\,
yz\left[2s-\hat{m}_c^2+\frac{s^2}{6}\delta\left(s-\hat{m}_{c}^{2}\right)\right]\nonumber\\
&&-\frac{m_s \langle\bar{s}g_{s}\sigma Gs\rangle}{128\pi^{4}}\int_{y_{i}}^{y_{f}}dy\,y\left(1-y\right)
\left(3s-2\tilde{m}_c^2\right)
-\frac{3m_s m_{c}^2\langle\bar{q}g_{s}\sigma Gq\rangle}{64\pi^{4}}\int_{y_{i}}^{y_{f}}dy\ ,
\end{eqnarray}

\begin{eqnarray}
\rho^{6}_{1}(s)&=&-\frac{m_{c}^{2}\langle\bar{q}q\rangle\langle\bar{s}s\rangle}{16\pi^{2}}\int_{y_{i}}^{y_{f}}dy
 +\frac{m_s m_c g_{s}^{2}\langle\bar{q}q\rangle^{2}}{288\pi^{4}}\int_{y_{i}}^{y_{f}}dy\int_{z_{i}}^{1-y}dz\
\left(\frac{1}{y}+\frac{1}{z}\right)  \nonumber\\
&&+\frac{g_{s}^{2}\left(\langle\bar{q}q\rangle^{2}+\langle\bar{s}s\rangle^{2}\right)}{288\pi^{4}}
\int_{y_{i}}^{y_{f}}dy\int_{z_{i}}^{1-y}dz\ yz
\left[2s-\hat{m}_{c}^{2}+\frac{s^2}{6}\delta\left(s-\hat{m}_{c}^{2}\right)\right] \nonumber\\
&&-\frac{g_{s}^{2}\left(\langle\bar{q}q\rangle^{2}+\langle\bar{s}s\rangle^{2}\right)}{288\pi^{4}}
\int_{y_{i}}^{y_{f}}dy\int_{z_{i}}^{1-y}dz\
\left(\frac{y}{z}+\frac{z}{y}\right)\left(1-y-z\right)\left(3s-2\hat{m}_{c}^{2}\right) \nonumber\\
&&-\frac{m_c^2g_{s}^{2}\left(\langle\bar{q}q\rangle^{2}+\langle\bar{s}s\rangle^{2}\right)}{576\pi^{4}}
\int_{y_{i}}^{y_{f}}dy\int_{z_{i}}^{1-y}dz\
\left(\frac{y}{z^2}+\frac{z}{y^2}\right)\left(1-y-z\right)
\left[2+s\delta(s-\hat{m}_{c}^{2})\right] \nonumber\\
&&-m_s m_c\left(\frac{\langle\bar{q}q\rangle\langle\bar{s}s\rangle}{64\pi^{2}}
+\frac{g_{s}^{2}\langle\bar{q}q\rangle^{2}}{3456\pi^{4}}\right) \int_{y_{i}}^{y_{f}}dy\
\left[2+s\delta \left(s-\tilde{m}_{c}^{2}\right)\right]
\nonumber\\
&&+\frac{m_s m_c^3 g_{s}^{2}\langle\bar{q}q\rangle^{2}}{576\pi^{4}}\int_{y_{i}}^{y_{f}}dy\int_{z_{i}}^{1-y}dz\
 \left(\frac{1}{y^2}+\frac{1}{z^2}\right)\delta(s-\hat{m}_{c}^{2})
\ ,
\end{eqnarray}

\begin{eqnarray}
\rho^{6}_{2}(s)&=&\frac{g_{s}^{2}\left(\langle\bar{q}q\rangle^2+\langle\bar{s}s\rangle^2\right)}{144\pi^{4}}
\int_{y_{i}}^{y_{f}}dy\int_{z_{i}}^{1-y}dz
\left[yz-2\left(y+z\right)\left(1-y-z\right)\right]
\left[2s-\hat{m}_{c}^{2}+\frac{s^2}{6}\delta(s-\hat{m}_{c}^{2})\right]
\nonumber\\
&& -\frac{m_{c}^{2}\langle\bar{q}q\rangle\langle\bar{s}s\rangle}{4\pi^{2}}\int_{y_{i}}^{y_{f}}dy
-m_s m_c\left(\frac{\langle\bar{q}q\rangle\langle\bar{s}s\rangle}{32\pi^{2}}
+\frac{g_{s}^{2}\langle\bar{q}q\rangle^{2}}{1728\pi^{4}}\right)
\int_{y_{i}}^{y_{f}}dy
\left[2+s\delta\left(s-\tilde{m}_c^2\right)\right] \nonumber\\
&&+\frac{g_{s}^{2}\left(\langle\bar{q}q\rangle^2+\langle\bar{s}s\rangle^2\right)}{864\pi^{4}}
\int_{y_{i}}^{y_{f}}dy\,
y\left(1-y\right)\left(3s-2\tilde{m}_c^2\right) \nonumber\\
&&-\frac{g_s^2\left(\langle\bar{q}q\rangle^2+\langle\bar{s}s\rangle^2\right)}{288\pi^{4}}
\int_{y_{i}}^{y_{f}}dy\int_{z_{i}}^{1-y}dz
\left(\frac{y}{z}+\frac{z}{y}\right)\left(1-y-z\right)(3s-2\hat{m}_{c}^{2})
\nonumber\\
&&-\frac{m_c^2g_s^2\left(\langle\bar{q}q\rangle^2+\langle\bar{s}s\rangle^2\right)}{864\pi^{4}}
\int_{y_{i}}^{y_{f}}dy\int_{z_{i}}^{1-y}dz
\left(\frac{y}{z^2}+\frac{z}{y^2}\right)\left(1-y-z\right)
\left[2+s\delta(s-\hat{m}_{c}^{2})\right]
\nonumber\\
&&+\frac{m_s m_{c}g_{s}^{2}\langle\bar{q}q\rangle^2}{288\pi^{4}}
\int_{y_{i}}^{y_{f}}dy\int_{z_{i}}^{1-y}dz\
\left[2\left(\frac{1}{y}+\frac{1}{z}\right)
+\left(\frac{1}{y^2}+\frac{1}{z^2}\right)m_c^2\delta(s-\hat{m}_{c}^{2})\right]
\ ,
\end{eqnarray}

\begin{eqnarray}
\rho^{7}_{1}(s)&=&\frac{m_{c}\left(\langle\bar{s}s\rangle-\langle\bar{q}q\rangle\right)}{256\pi^{2}}
\langle\frac{\alpha_{s}GG}{\pi}\rangle \int_{y_{i}}^{y_{f}}dy\int_{z_{i}}^{1-y}dz
\left[\left(\frac{y}{z^2}+\frac{z}{y^2}\right)\left(1-y-z\right)+2\right]
 \left[2+s\delta\left(s-\hat{m}_{c}^{2}\right)\right] \nonumber\\
&&+\frac{m_{c}^3\left(\langle\bar{q}q\rangle-\langle\bar{s}s\rangle\right)}{768\pi^{2}}
\langle\frac{\alpha_{s}GG}{\pi}\rangle
\int_{y_{i}}^{y_{f}}dy\int_{z_{i}}^{1-y}dz
\left(\frac{y}{z^3}+\frac{z}{y^3}+\frac{1}{y^2}+\frac{1}{z^2}\right)
\left(1-y-z\right) \nonumber\\
&&\left(1+\frac{s}{T^2}\right)
\delta\left(s-\hat{m}_{c}^{2}\right)
+\frac{m_{c}\left(\langle\bar{s}s\rangle-\langle\bar{q}q\rangle\right)}{1536\pi^{2}}
\int_{y_{i}}^{y_{f}}dy \left[2+s\delta\left(s-\tilde{m}_{c}^{2}\right)\right]
\nonumber\\
&&+\frac{m_s m_{c}^{2}\langle\bar{q}q\rangle}{128\pi^{2}}\langle\frac{\alpha_{s}GG}{\pi}\rangle
\int_{y_{i}}^{y_{f}}dy\int_{z_{i}}^{1-y}dz
\left(\frac{1}{y^2}+\frac{1}{z^2}\right)\delta(s-\hat{m}_{c}^{2})  \nonumber\\
&&-\frac{m_s m_{c}^2\langle\bar{s}s\rangle}{384\pi^{2}}\langle\frac{\alpha_{s}GG}{\pi}
\rangle\int_{y_{i}}^{y_{f}}dy\int_{z_{i}}^{1-y}dz
\left(\frac{y}{z^2}+\frac{z}{y^2}\right)\left(1-y-z\right)
\left[1+\frac{s}{T^2}+\frac{s^2}{2T^4}\right]
\delta(s-\hat{m}_{c}^{2}) \nonumber\\
&&-\frac{m_s m_{c}^4\langle\bar{q}q\rangle}{384\pi^{2}T^2}\langle\frac{\alpha_{s}GG}{\pi}\rangle
\int_{y_{i}}^{y_{f}}dy\int_{z_{i}}^{1-y}dz
\left(\frac{1}{y^3}+\frac{1}{z^3}\right)\delta(s-\hat{m}_{c}^{2}) \nonumber\\
&&+\frac{m_s\langle\bar{s}s\rangle}{256\pi^{2}}\langle\frac{\alpha_{s}GG}{\pi}\rangle
\int_{y_{i}}^{y_{f}}dy\int_{z_{i}}^{1-y}dz\left(y+z\right)
\left[3+\left(2s+\frac{s^2}{2T^2}\right)\delta(s-\hat{m}_{c}^{2})\right] \nonumber\\
&&+\frac{m_s m_c^2\langle\bar{q}q\rangle}{768\pi^2}\langle\frac{\alpha_{s}GG}{\pi}\rangle
\int_{y_{i}}^{y_{f}}dy\left(1+\frac{s}{T^2}\right)\delta(s-\tilde{m}_{c}^{2})
\ ,
\end{eqnarray}

\begin{eqnarray}
\rho^{7}_{2}(s)&=&\frac{m_{c}\left(\langle\bar{q}q\rangle-\langle\bar{s}s\rangle\right)}{128\pi^{2}}
\langle\frac{\alpha_{s}GG}{\pi}\rangle\int_{y_{i}}^{y_{f}}dy\int_{z_{i}}^{1-y}dz \left[2-\left(\frac{y}{z^2}+\frac{z}{y^2}\right)\left(1-y-z\right)\right]
\left[2+s\delta(s-\hat{m}_{c}^{2})\right]
 \nonumber\\
&&+\frac{m_{c}^3\left(\langle\bar{q}q\rangle-\langle\bar{s}s\rangle\right)}{384\pi^{2}}
\langle\frac{\alpha_{s}GG}{\pi}\rangle
\int_{y_{i}}^{y_{f}}dy\int_{z_{i}}^{1-y}dz
\left(\frac{1}{y^{2}}+\frac{1}{z^{2}}+\frac{y}{z^{3}}+\frac{z}{y^{3}}\right)\left(1-y-z\right) \nonumber\\
&&\left(1+\frac{s}{T^2}\right)\delta(s-\hat{m}_{c}^{2})
-\frac{m_c\langle\bar{q}q\rangle-\left(m_c+m_s\right)\langle\bar{s}s\rangle}{768\pi^{2}}
\langle\frac{\alpha_{s}GG}{\pi}\rangle\int_{y_{i}}^{y_{f}}dy
\left[2+s\delta(s-\tilde{m}_{c}^{2})\right]
\nonumber\\
&&+\frac{m_s m_c^2\langle\bar{q}q\rangle}{96\pi^{2}}
\langle\frac{\alpha_{s}GG}{\pi}\rangle
\int_{y_{i}}^{y_{f}}dy\int_{z_{i}}^{1-y}dz
\left[3\left(\frac{1}{y^2}+\frac{1}{z^2}\right)
-\left(\frac{1}{y^3}+\frac{1}{z^3}\right)\frac{m_c^2}{T^2}\right]
\delta(s-\hat{m}_{c}^{2})\nonumber\\
&&-\frac{m_s m_c^2\langle\bar{s}s\rangle}{384\pi^{2}}
\langle\frac{\alpha_{s}GG}{\pi}\rangle
\int_{y_{i}}^{y_{f}}dy\int_{z_{i}}^{1-y}dz
\left(\frac{y}{z^2}+\frac{z}{y^2}\right)
\left(1+\frac{s}{T^2}\right)\delta(s-\hat{m}_{c}^{2})
\nonumber\\
&&-\frac{m_s m_c^2\langle\bar{s}s\rangle}{192\pi^{2}}
\langle\frac{\alpha_{s}GG}{\pi}\rangle
\int_{y_{i}}^{y_{f}}dy\int_{z_{i}}^{1-y}dz
\left(\frac{y}{z^2}+\frac{z}{y^2}\right)\left(1-y-z\right)
\left(1+\frac{s}{T^2}+\frac{s^2}{2T^4}\right)\delta(s-\hat{m}_{c}^{2})
\nonumber\\
&&-\frac{m_s\langle\bar{s}s\rangle}{384\pi^{2}}
\langle\frac{\alpha_{s}GG}{\pi}\rangle
\int_{y_{i}}^{y_{f}}dy\int_{z_{i}}^{1-y}dz
\left(y+z\right)
\left[3+\left(2s+\frac{s^2}{2T^2}\right)\delta(s-\hat{m}_{c}^{2})\right]
\nonumber\\
&&+\frac{m_s m_{c}^2\langle\bar{q}q\rangle}{192\pi^{2}}
\langle\frac{\alpha_{s}GG}{\pi}\rangle\int_{y_{i}}^{y_{f}}dy
\left(1+\frac{s}{T^2}\right)\delta(s-\tilde{m}_{c}^{2})
\ ,
\end{eqnarray}

\begin{eqnarray}
\rho^{8}_{1}(s)&=&\frac{\left(\langle\bar{q}q\rangle\langle\bar{s}g_{s}\sigma Gs\rangle
+\langle\bar{s}s\rangle\langle\bar{q}g_{s}\sigma Gq\rangle\right)
}{64\pi^{2}}\int_{y_{i}}^{y_{f}}dy
\left[m_c^2\left(1+\frac{s}{T^2}\right)-s\right]
\delta(s-\tilde{m}_{c}^{2}) \nonumber\\
&&+\frac{m_s m_c\left(2\langle\bar{q}q\rangle\langle\bar{s}g_{s}\sigma Gs\rangle
+3\langle\bar{s}s\rangle\langle\bar{q}g_{s}\sigma Gq\rangle\right)}{384\pi^{2}}\int_{y_{i}}^{y_{f}}dy\
\left(1+\frac{s}{T^2}+\frac{s^2}{2T^4}\right)\delta(s-\tilde{m}_{c}^{2}) \nonumber\\
&&+\frac{m_s m_c\langle\bar{s}s\rangle\langle\bar{q}g_{s}\sigma Gq\rangle}{64\pi^{2}}\int_{y_{i}}^{y_{f}}dy\
\left[1-\frac{1}{2y\left(1-y\right)}\right]\left(1+\frac{s}{T^2}\right)\delta(s-\tilde{m}_{c}^{2})
\ ,
\end{eqnarray}

\begin{eqnarray}
\rho^{8}_{2}(s)&=&\frac{m_s m_c\left(2\langle\bar{q}q\rangle\langle\bar{s}g_{s}\sigma Gs\rangle
+3\langle\bar{s}s\rangle\langle\bar{q}g_{s}\sigma Gq\rangle\right)}{192\pi^{2}}
\int_{y_{i}}^{y_{f}}dy\
\left(1+\frac{s}{T^2}+\frac{s^2}{2T^4}\right)\delta(s-\tilde{m}_{c}^{2})\nonumber\\
&&+\frac{m_c^2\left(\langle\bar{q}q\rangle\langle\bar{s}g_{s}\sigma Gs\rangle
+\langle\bar{s}s\rangle\langle\bar{q}g_{s}\sigma Gq\rangle\right)}{16\pi^{2}}\int_{y_{i}}^{y_{f}}dy
\left(1+\frac{s}{T^2}\right)\delta(s-\tilde{m}_{c}^{2}) \ ,
\end{eqnarray}

\begin{eqnarray}
\rho^{10}_{1}(s)&=&-\frac{\langle\bar{q}g_{s}\sigma Gq\rangle\langle\bar{s}g_{s}\sigma Gs\rangle}{64\pi^{2}}
\int_{y_{i}}^{y_{f}}dy\,
\left(\frac{49s}{48T^2}-\frac{s^2}{2T^4}+\frac{m_c^2s^2}{4T^6}\right) \delta(s-\tilde{m}_{c}^{2}) \nonumber\\
&&+\frac{\langle\bar{q}q\rangle\langle\bar{s}s\rangle}{288}\langle\frac{\alpha_{s}GG}{\pi}\rangle
\int_{y_{i}}^{y_{f}}dy\
\left\{\left(\frac{1}{y^3}+\frac{1}{\left(1-y\right)^3}\right)\frac{m_c^4}{T^4}-\frac{m_c^2s^2}{T^6} \right. \nonumber\\
&&\left.-3\left[\frac{1}{y^2}+\frac{1}{\left(1-y\right)^2}\right]\frac{m_c^2}{T^2}\right\}\delta(s-\tilde{m}_{c}^{2}) \nonumber\\
&&-\frac{m_s m_{c}\langle\bar{q}q\rangle\langle\bar{s}s\rangle}{384}\langle\frac{\alpha_{s}GG}{\pi}\rangle\int_{y_{i}}^{y_{f}}dy\ \left\{\left[\frac{1-y}{y^2}+\frac{y}{\left(1-y\right)^2}\right]\frac{s}{T^4}+\frac{s^3}{6T^8}\right\}
 \delta(s-\tilde{m}_{c}^{2}) \nonumber\\
&&+\frac{m_s m_c\langle\bar{q}g_{s}\sigma Gq\rangle\langle\bar{s}g_{s}\sigma Gs\rangle}{768\pi^{2}}
\int_{y_{i}}^{y_{f}}dy\,
\left\{\left[\frac{1}{y\left(1-y\right)}-2\right]\frac{s^2}{T^6}-\frac{s^3}{2T^8}\right\} \delta(s-\tilde{m}_{c}^{2}) \nonumber\\
&&-\frac{m_s m_{c}^{3}\langle\bar{q}q\rangle\langle\bar{s}s\rangle}{1152T^4}\langle\frac{\alpha_{s}GG}{\pi}\rangle\int_{y_{i}}^{y_{f}}dy\ \left[\frac{1}{y^3}+\frac{1}{\left(1-y\right)^3}\right]
\left(1-\frac{s}{T^2}\right) \delta(s-\tilde{m}_{c}^{2})
 \ ,
\end{eqnarray}

\begin{eqnarray}
\rho^{10}_{2}(s)&=&-\frac{m_{c}^{2}\langle\bar{q}g_{s}\sigma Gq\rangle\langle\bar{s}g_{s}\sigma Gs\rangle}{64\pi^{2}T^6}
\int_{y_{i}}^{y_{f}}dy\
s^2 \delta(s-\tilde{m}_{c}^{2})
-\frac{\langle\bar{q}g_{s}\sigma Gq\rangle\langle\bar{s}g_{s}\sigma Gs\rangle}{256\pi^{2}T^2}
\int_{y_{i}}^{y_{f}}dy\
s \delta(s-\tilde{m}_{c}^{2})
\nonumber\\
&&-\frac{m_s m_{c}\langle\bar{q}g_{s}\sigma Gq\rangle\langle\bar{s}g_{s}\sigma Gs\rangle}{768\pi^{2}T^8}
\int_{y_{i}}^{y_{f}}dy\
s^3 \delta(s-\tilde{m}_{c}^{2})\nonumber\\
&&+\frac{m_c^2\langle\bar{q}q\rangle\langle\bar{s}s\rangle}{72T^2}
\langle\frac{\alpha_{s}GG}{\pi}\rangle
\int_{y_{i}}^{y_{f}}dy
\left\{\left[\frac{1}{y^3}+\frac{1}{(1-y)^3}\right]\frac{m_c^2}{T^2}
-3\left[\frac{1}{y^2}+\frac{1}{(1-y)^2}\right]\right\}
\delta(s-\tilde{m}_{c}^{2})
\nonumber\\
&&-\frac{m_{c}^{2}\langle\bar{q}q\rangle\langle\bar{s}s\rangle}{72T^6}
\langle\frac{\alpha_{s}GG}{\pi}\rangle
\int_{y_{i}}^{y_{f}}dy\,
s^2 \delta(s-\tilde{m}_{c}^{2}) \nonumber\\
&&-\frac{m_s m_{c}\langle\bar{q}q\rangle\langle\bar{s}s\rangle}{1152T^4}
\langle\frac{\alpha_{s}GG}{\pi}\rangle\int_{y_{i}}^{y_{f}}dy\
\left\{\frac{s^3}{T^4}
+6s\left[\frac{1}{y^2}+\frac{1}{(1-y)^2}-\frac{1}{y(1-y)}\right]\right\}
\delta(s-\tilde{m}_{c}^{2}) \nonumber\\
&&-\frac{m_s m_{c}^{3}\langle\bar{q}q\rangle\langle\bar{s}s\rangle}{576T^4}
\langle\frac{\alpha_{s}GG}{\pi}\rangle
\int_{y_{i}}^{y_{f}}dy\
\left[\frac{1}{y^3}+\frac{1}{\left(1-y\right)^3}\right]
\left(1-\frac{s}{T^2}\right) \delta(s-\tilde{m}_{c}^{2})\ ,
\end{eqnarray}
where $y_{f}=\frac{1+\sqrt{1-4m_{c}^{2}/s}}{2}$, $y_{i}=\frac{1-\sqrt{1-4m_{c}^{2}/s}}{2}$, $z_{i}=\frac{ym_{c}^{2}}{ys-m_{c}^{2}}$, $\hat{m}_{c}^{2}=\frac{(y+z)m_{c}^{2}}{yz}$, $\tilde{m}_{c}^{2}=\frac{m_{c}^{2}}{y(1-y)}$, $\int_{y_{i}}^{y_{f}}dy\rightarrow\int_{0}^{1}$, $\int_{z_{i}}^{1-y}dz\rightarrow\int_{0}^{1-y}dz$, when the $\delta$ functions $\delta(s-\hat{m}_{c}^{2})$ and $\delta(s-\tilde{m}_{c}^{2})$ appear.

\end{document}